\documentclass[fleqn,usenatbib]{mnras}

\usepackage{mathptmx}

\usepackage[T1]{fontenc}

\DeclareRobustCommand{\VAN}[3]{#2}
\let\VANthebibliography\thebibliography
\def\thebibliography{\DeclareRobustCommand{\VAN}[3]{##3}\VANthebibliography}

\usepackage{graphicx}	
\usepackage{amsmath}	
\usepackage{amssymb}	

\usepackage{floatrow}
\usepackage{subcaption}
\usepackage{caption}
\usepackage{hyperref}
\hypersetup{
    colorlinks=true,
    linkcolor=blue,
    filecolor=green,
    urlcolor=blue,
}
\usepackage{multicol}
\usepackage{multirow}
\usepackage{xcolor}
\usepackage{todonotes}
\usepackage[markup=underlined]{changes}
\setaddedmarkup{\textbf{\textcolor{black}{#1}}}

\definecolor{MBlue}{RGB}{001, 031, 191}
\definecolor{MGreen}{RGB}{033, 087, 050}

\title[Location of GeV/Optical Outbursts in Blazars]{Location of a Sample of GeV and Optical Outbursts in the Jets of Blazars}

\author[Kundu et al.]{
Maitreya Kundu$^{1, 2}$\thanks{Email --- \href{mailto:k.maitreya@wustl.edu}{\texttt{k.maitreya@wustl.edu}}},
Arit Bala$^{2}$,
Saugata Barat$^{3, 4}$,
and Ritaban Chatterjee$^{2}$ \\
\\
$^{1}$Department of Physics, Washington University in St.\ Louis, 1 Brookings Drive, St.\ Louis, MO 63130, USA \\
$^{2}$School of Astrophysics, Presidency University, 86/1 College Street, Kolkata-700073, India \\
$^{3}$Anton Pannekoek Institute for Astronomy, University of Amsterdam, Science Park 904, NL-1098 XH Amsterdam, the Netherlands \\ 
$^{4}$Kavli Institute for Astrophysics and Space Research, Massachusetts Institute of Technology, Cambridge, MA 02139, USA
}

\date{Accepted XXX. Received YYY; in original form ZZZ}

\pubyear{2023}

\begin{document}
\label{firstpage}
\pagerange{\pageref{firstpage}--\pageref{lastpage}}
\maketitle

\begin{abstract}
The exact location of the $\gamma$-ray emitting region in blazar jets has long been a matter of debate. However, the location has important implications about the emission processes, geometric and physical parameters of the jet, as well as the nature of interaction of the jet with the interstellar and intergalactic medium. Diverse conclusions have been drawn by various authors based on a variety of methods applied to different data sets of many blazars, \textit{e.g.}, the location is less than 0.1 pc from the central engine within the broad line region (BLR) or a few or tens of pc downstream beyond the dusty torus or at some intermediate distance. Here we use a method, established in a previous work, in which the location of the GeV/optical emission is determined using the ratio of energy dissipated during contemporaneous outbursts at those wave bands. We apply it to a total of 47 multi-wavelength outbursts in 10 blazars. We find that the location of the GeV/optical emission is beyond the BLR for all cases. This result is consistent with other studies, in which the location has been determined for a large sample of blazars. We compare the location determined by our method for several GeV outbursts of multiple blazars to that obtained by other authors using different methods. 
We find that our results are consistent in such one-to-one comparison {in most cases}, for which the required data were available.
\end{abstract}




\section{Introduction}

Blazars are a particular type of jetted active galactic nuclei (AGNs) with their jet at a very close angle to, or pointing directly at, the observers \citep{UrryPadovani1995, BlandfordRees1978}. Thus, due to relativistic beaming, the emission from the jet dominates that from other AGN components, \textit{e.g.}, the accretion disk, broad line region (BLR), and dusty torus (DT). The amplified emission from the jet offers us a way to probe its inner workings --- emission processes, particle populations, magnetic field, etc. --- in much more detail than is possible for other classes of AGNs \citep[see \textit{e.g.},][for a review]{Hovatta2019, Blandford2019ARAA}.

Due to the observed spectral nature and high polarization fraction, it is believed that the radio to optical (sometimes extending to X-ray energies) emission in blazar jets is due to synchrotron radiation by relativistic electrons \citep{Bregman1981, UrryMushotzky1982, Impey1988, Marscher1998IAU}.
In the `leptonic model' of blazar emission, the $\gamma$-rays in the jet are produced by the inverse-Compton scattering of those synchrotron photons \citep[\textit{e.g.},][]{Maraschi1992, Sikora1994, bott07} or photons from the BLR or DT \citep[\textit{e.g.},][]{DermerSchlickeiser1993, ghise1998, blejo2000, bott10}. IC emission through the above two kinds of seed photons are called synchrotron self-Compton (SSC) and external Compton (EC) process, respectively. However, determination of the exact location of the $\gamma$-ray emission region within the jet is still an open problem. It may be closer to the black hole, \textit{e.g.}, 0.1 pc or nearer, \textit{i.e.}, within the BLR or farther down the jet at a few or tens of pc, \textit{i.e.}, away from the BLR, near or beyond the torus.

During the \textit{Fermi} era many authors have studied various observed properties of blazars in order to constrain the location of GeV emission. However, the results have not converged. On the one hand, it has been argued, based on spectral energy distribution (SED) modeling, that strong GeV emission can only be produced if large amount of seed photons from the BLR is supplied, for which the emission region needs to be within the BLR \citep[\textit{e.g.},][]{Sikora1994, Wendel2021}. On the other hand, the emission region is supposed to be outside the BLR in blazars flaring at TeV energies along with GeV emission because otherwise the TeV photons will be partially or fully absorbed by the BLR due to the $\gamma$-$\gamma$ absorption process \citep{Costamante2018,Tavecchio2009, Donea2003}.

In addition, various studies have found differing results. Location within the BLR has been inferred by comparing the GeV variability over a decade with the radio emission at 15 GHz of a sample of joint MOJAVE-\textit{Fermi} monitored AGNs \citep{Kramarenko2022, Kang2021}; using X-ray features at different flux-states of the blazar PKS 2005-489 \citep{Chase2023}; and from the lack of observed cut-off in their GeV spectra in some blazars \citep{Acharyya2021, Agarwal2024}. The broadband SED of many blazars have been satisfactorily fit by models in which the GeV emission is assumed to be within the BLR \citep[\textit{e.g.},][]{Ghisellini2010}.

On the other hand, location outside of the BLR, near or even beyond the torus have been obtained for several blazars by observing the temporal lag between $\gamma$-ray flares and the movement of radio bright knots, monitored by the Very Long Baseline Array, along the jet \citep{Marscher2008Nature, Marscher2010, Marscher2011JApA, Agudo2011, Jorstad2013, Rani2014, Marscher2016, Troitskiy2016, Acciari2022, Nalewajko2014}, a systematic study of the Compton dominance in more than 60 blazars \citep{Harvey2020}, from the absence of any observable signature of the Klein-Nishina effect, \textit{e.g.}, at the GeV spectra of 21 blazars \citep{Cao2013}, using temporal variability at day to sub-day timescales in several blazars \citep{Dotson2015, Saito2013}, and by the identification of the seed photon population being up-scattered in the EC process to be infrared and hence originating in the DT in a sample of flat spectrum radio quasars \citep{Arsioli2018}. Finally, in some cases, authors have found evidence that the location of multiple GeV flares may be significantly different from each other, some being within the BLR and the rest beyond it \citep[\textit{e.g.},][]{Brown2013, Prince2019}. \citet{Patino-Alvarez2018} have suggested that the location of the $\gamma$-ray emitting region is dependent on the activity state of the source, and may change when transitioning from a flaring to a non-flaring state.

In the absence of a clear result or pattern found in the various investigation of the location of the GeV emission in blazar jets using different methods, extending the analyses to larger samples and comparison among results obtained by multiple methods is necessary. \citet{Barat2022} established a method to infer the location by using the ratio of dissipated energy of contemporaneous outbursts at the GeV and optical bands, and its comparison with multi-wavelength outbursts in simulated light curves of nonthermal emission in blazars. While \citet{Barat2022} identified the contemporaneous outbursts at the GeV and optical bands and determined their optical to $\gamma$-ray energy dissipation ratio in a sample of blazars similar to that used here, they actually found the constraints on the location of the emission region for only two cases in order to demonstrate the method. In this work, we apply that method to our full sample of blazar flares. In addition to obtaining the location for more blazar flares using our method, another goal of this work is to compare our results with that obtained by other methods for the exact same outbursts. For several sources in our sample, the location of the GeV emitting region has been inferred using other methods. For example, \citet{Rani2018} have carried out a detailed study of the 2013--2014 flaring state of 3C 279 to constrain the location. They suggested the existence of two emission zones, one upstream and one further down the jet, which produced GeV emission at the passing of two shock fronts.
Other studies such as \citet{Coogan2016, Khatoon2022, Pacciani2014, Hayashida2015, Lisakov2017} have also probed the GeV emission zones of different blazars in our sample using SEDs or variability/cooling timescales. We intend to analyze the GeV/optical emission of those blazars during those exact epochs and and compare our results with those found by other authors.

This paper is organized as follows. In the following section (\S\ref{sec:data}), we describe the process of optical ($R$ band) and $\gamma$-ray data acquisition. In \S\ref{sec:method}, we describe our formalism and methodology for finding the GeV emission region from the optical and $\gamma$-ray light curves, along with the numerical simulation of the jet emission. In \S\ref{sec:results}, the application of our formalism in finding the $\gamma$-ray emitting regions of several \textit{Fermi} blazars is discussed. We also compare our results with that found by several pre-existing studies. The existence of two different classes of outbursts is also highlighted. In \S\ref{sec:discussion}, we summarize our results and point out several details and nuances of our method. 

We note that the data reduction and methodology we use in this work is similar to that described in \citet{Barat2022}. However, we briefly discuss those here for easy reference to the readers.

\section{Data for Multi-wavelength Outbursts}
\label{sec:data}

In most low-synchrotron peaked (LSP) blazars, having the frequency of the synchrotron peak below $10^{14}$ Hz \citep{PadovaniGiommi1995, Abdo2010b}, the $0.1-300$ GeV emission is dominated by the EC process with seed photons supplied by the BLR and/or the DT. The GeV and optical variability at weeks to months timescale are strongly correlated with small or zero time delay in those blazars \citep[\textit{e.g.},][]{Zhang2021, Majumdar2019, Isler2015, Liodakis2019}. Those sources are relevant for this work. In many high synchrotron peaked (HSP) blazars, having the frequency of the synchrotron peak above $10^{15}$ Hz, the GeV emission is due to the SSC process \citep[e.g.,][]{Chiang2002, MAGIC2018, Zhang2013}. Some authors have concluded that in certain blazar outbursts the seed photon for the IC process may be supplied by the part of the jet away from the axis, called the sheath while the jet axis contains the highest energy electrons which scatter those photons \citep{Ghisellini2005, MacDonald2017, Giroletti2004}. Several authors have shown that the higher energy emission in blazar jets may be explained in the `hadronic scenario,' in which the $\gamma$-ray emission may be due to proton synchrotron radiation, proton-initiated cascades and the interaction of photons with the secondary particles \citep[see, \textit{e.g.},][]{Mannheim1993, Mucke2001, MuckeProtheroe2003, Ackermann2016, Bottacini2016}. This work is not applicable to those blazars and/or in those context.


Therefore, in this work, we consider a sample of low synchrotron-peaked (LSP) blazars. The $\gamma$-ray light curves are drawn {from \textit{Fermi} Large Area Telescope \citep[\textit{Fermi}-LAT;][]{FermiLAT2009}}, and the optical data are retrieved from the Yale-SMARTS blazar monitoring program. We select sources, which have been monitored at the optical wavelengths during 2008--16 and have undergone more than ten outbursts at the $\gamma$-ray energies during those intervals. We consider an increase in flux to be an outburst or a flare if the said increase is at least $3 \sigma$ above the long term average flux level of the blazar, where $\sigma$ is the standard deviation about the average flux level. We have only chosen sources which have continuous coverage (at least one observation in each monthly time-bin) in both $\gamma$-ray and optical bands. Additionally, these sources display contemporaneous outbursts in both bands, and each source has at least one such pair of flares. We have selected a total of ten blazars in our sample. In addition, we also analyze certain flare-pairs in the blazar 3C 279 for direct comparison with results found by other authors.

\begin{table*}
\begin{center}
\begin{tabular}{c|c|c|c|c}
\hline
\hline
 Object Name & Redshift & \textit{Fermi} time interval [MJD] & SMARTS time interval [MJD] \\
\hline
\hline
PKS 0208-512 & 1.003 & 54721--58509 & 54501--57060 \\
{PKS 0402-362} & {1.423} & {54908--57615} & {55838--57297} \\
{PKS 0426-380} & {1.112} & {54683--57621} & {56282--57144} \\
PKS 0454-234 & 1.002 & 54690--57344 & 55861--57145 \\
3C 273  & 0.158 & 54683--57860 & 54537--56791 \\
PKS 1244-255  & 0.638 & 54693--57955 & 55676--57115 \\
3C 279 & 0.536 & 54683--57450 &  54501--57404 \\
PKS 1510-089 & 0.361 & 54697--57820 & 54501--57204 \\
3C 454.3 & 0.859 & 54683--57064 & 54640--57356 \\
{PKS 2142-75} & {1.139} & {55183--58172} & {55297--56999} \\
PKS 2326-502 & 0.518 & 54917--57077 &  56109--57201 \\
\hline
\end{tabular}
\end{center}
\caption{Our full sample of LSP blazars. All of these sources have well sampled optical and $\gamma$-ray light curves. See text for sample selection criteria.}
\label{tab:sample}
\end{table*}

\subsection{GeV Data}

{The $\gamma$-ray data of the blazars in our sample are retrieved from the FSSC FTP server\footnote{{\url{https://heasarc.gsfc.nasa.gov/FTP/fermi/data/lat/}}}. We use the weekly all-sky data files, which are suitable for extracting the light curves of a larger number of sources. We employ the {\tt Fermitools} (version {\tt v11r5p3}) software and the \texttt{P8R2\_SOURCE\_V6} instrument response functions to carry out unbinned likelihood analysis of the data. We select the data for sources in our sample listed in Table \ref{tab:sample} and the time ranges given there, and using the energy range 0.1-100 GeV. We have used the \texttt{P8R3\_SOURCE} class photons, a zenith angle cut of 90$^{\circ}$, and exclude photons from regions in the sky and time intervals containing known GRBs and solar flares. A region of interest (RoI) of $12^{\circ}$ centered at the target source is used. In the unbinned likelihood analysis, we include all sources from the 4FGL-DR3 catalog of \textit{Fermi}-detected sources \citep{Abdollahi2020_4FGL, Abdollahi2022} within a radius of 30$^{\circ}$ centered at the RoI. The target blazars are modeled with a power-law, with the normalization and spectral index kept free. For other variable sources in the region, the spectral parameters were kept frozen at catalog value and normalization was free. The model also includes Galactic and isotropic background components {\tt gll\_iem\_v07} and {\tt iso\_P8R3\_SOURCE\_V3\_v1}\footnote{These models are available at \url{http://fermi.gsfc.nasa.gov/ssc/data/access/lat/BackgroundModels.html}}, respectively. Sources with test-statistic (TS) value \citep{Mattox1996} $\le 25$ were rejected as insignificant detections.}

\subsection{Optical data}

We have used optical ($R$-band) data from the Yale-SMARTS Blazar Monitoring Program\footnote{\url{http://www.astro.yale.edu/smarts/glast/home.php}}, which monitored the \textit{Fermi}-observed blazars. The observations were carried out with the ANDICAM instrument on the SMARTS 1.3 m optical telescope in CTIO, Chile, which performed $BVRJK$ photometric observations with a variable cadence for different sources. \citet{Bonning2012} describe in detail the procedure of data acquisition, reduction, and analysis.

\section{Methodology}
\label{sec:method}

To constrain the $\gamma$-ray emitting region of the blazars in our sample, we use the formalism of \citet{Barat2022}. We extract near-simultaneous flares in the optical and $\gamma$-ray bands and estimate the ratio of energies emitted in these two bands for each such pair. We use a numerical simulation of the emission from the jet, and for the simulated light curves we perform the same analysis. For the light curves simulated with the emission region kept at different distances from the central black hole, we compare the results with that obtained from the observed data, to draw our conclusions regarding the distance of the emitting region along the jet.

\subsection{Analysis of the observed data}
\label{subsec:analysisObs}


To determine the energies of individual optical or $\gamma$-ray flares, first we decompose a raw light curve into distinct double exponential flares, following the prescription of \citet{Valtaoja1999, Chatterjee2012}. Each flare is modeled using the double exponential function (representing the rapid rise and fall of intensity):
\begin{equation}\label{eqn:flare}
    f(t) =
    \begin{cases}
        A_0 \exp\left(\frac{t - t_0}{\tau_{\rm rise}}\right) & \text{if } t < t_0 \\
        A_0 \exp\left(\frac{t_0 - t}{\tau_{\rm fall}}\right) & \text{if } t > t_0
    \end{cases}
\end{equation}
where, $t_0$ is the time when the flare reaches its maximum intensity. The quantities $\tau_{\rm rise}$ and $\tau_{\rm fall}$ represent the rise and decay timescales of each flare, respectively. They are free parameters to the fitting function (Equation \ref{eqn:flare}), along with $t_0$ and $A_0$, and are determined by maximum likelihood fitting to a raw light curve. $A_0$ is a normalization.

In the above decomposition, we use light curves smoothed using a Gaussian function of width 10 days. Blazar variability is typically of red noise nature, \textit{i.e.}, longer-term fluctuations have higher amplitude and therefore we are interested in the more prominent long-term outbursts.

\begin{figure*}
    \begin{center}
        \includegraphics[scale = 0.23]{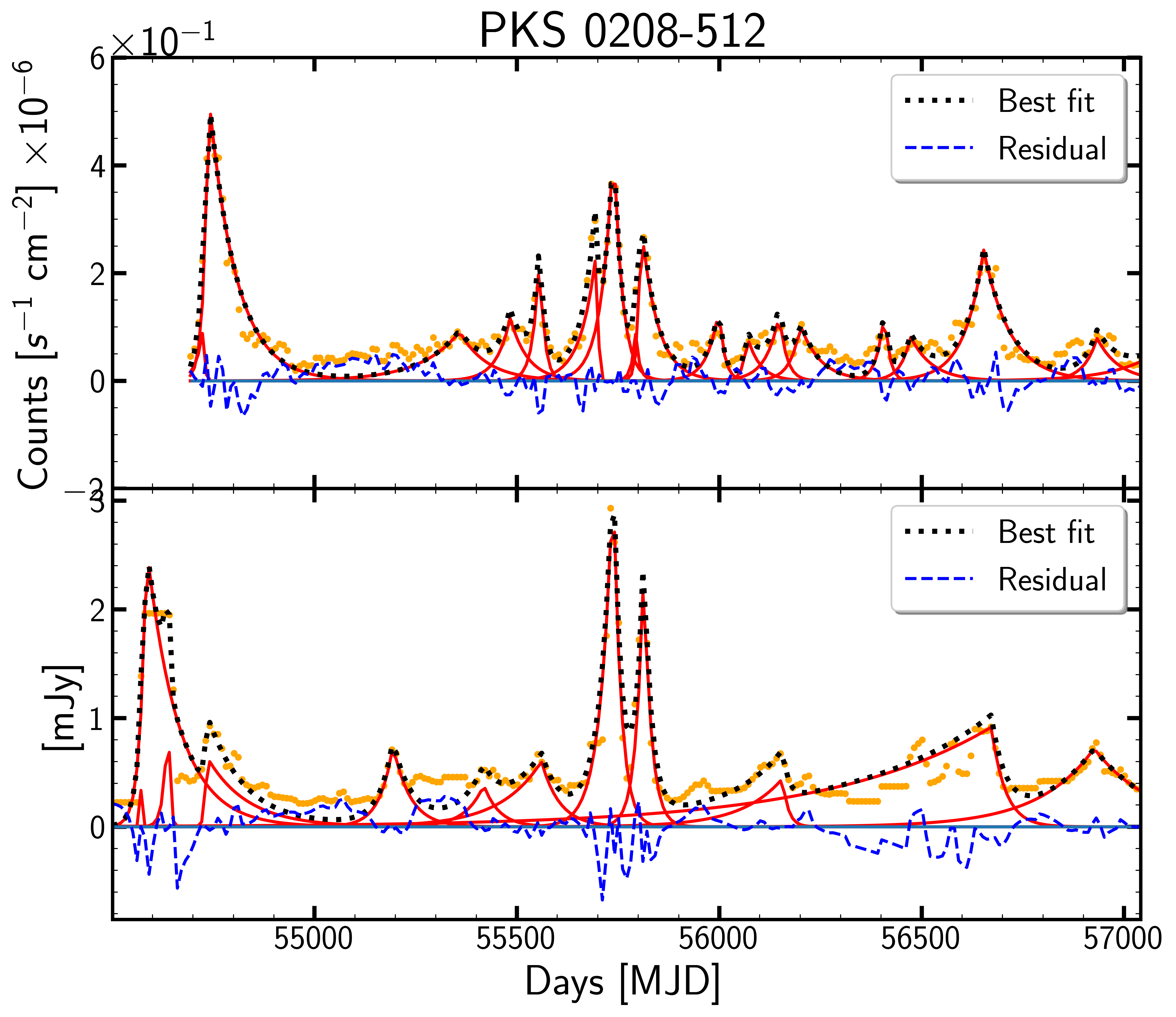}
        \includegraphics[scale = 0.23]{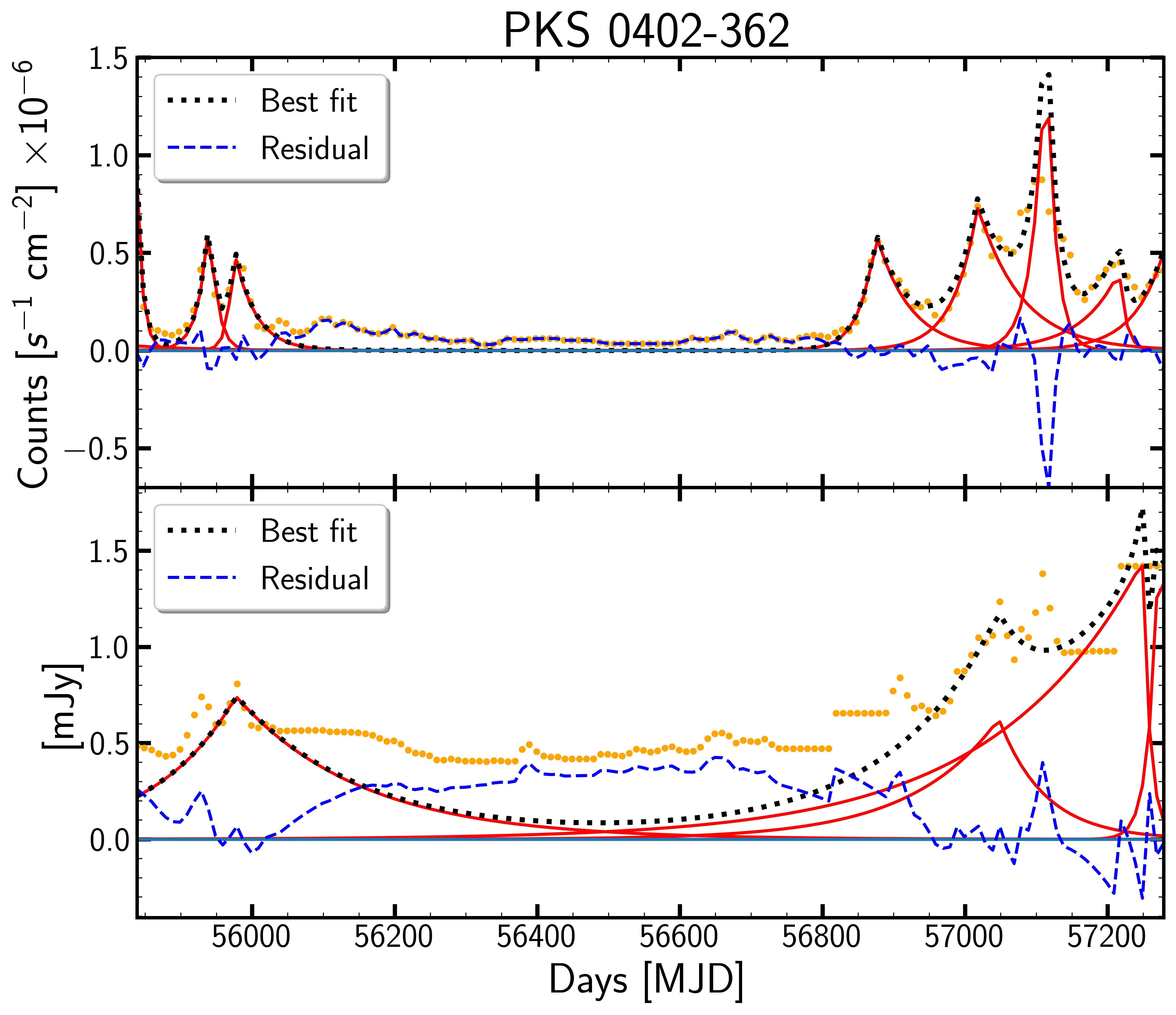}
        \includegraphics[scale = 0.23]{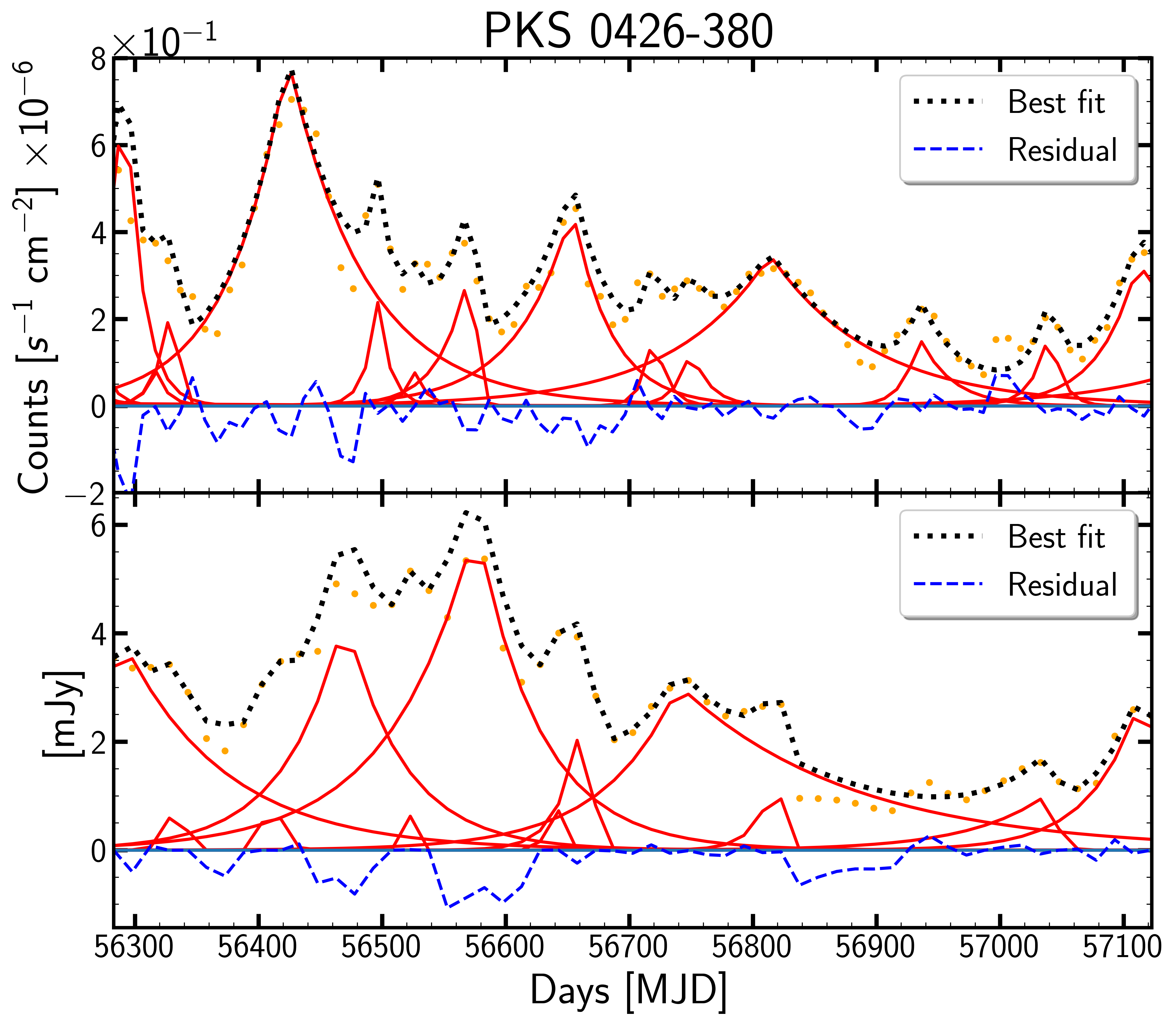}
        \includegraphics[scale = 0.23]{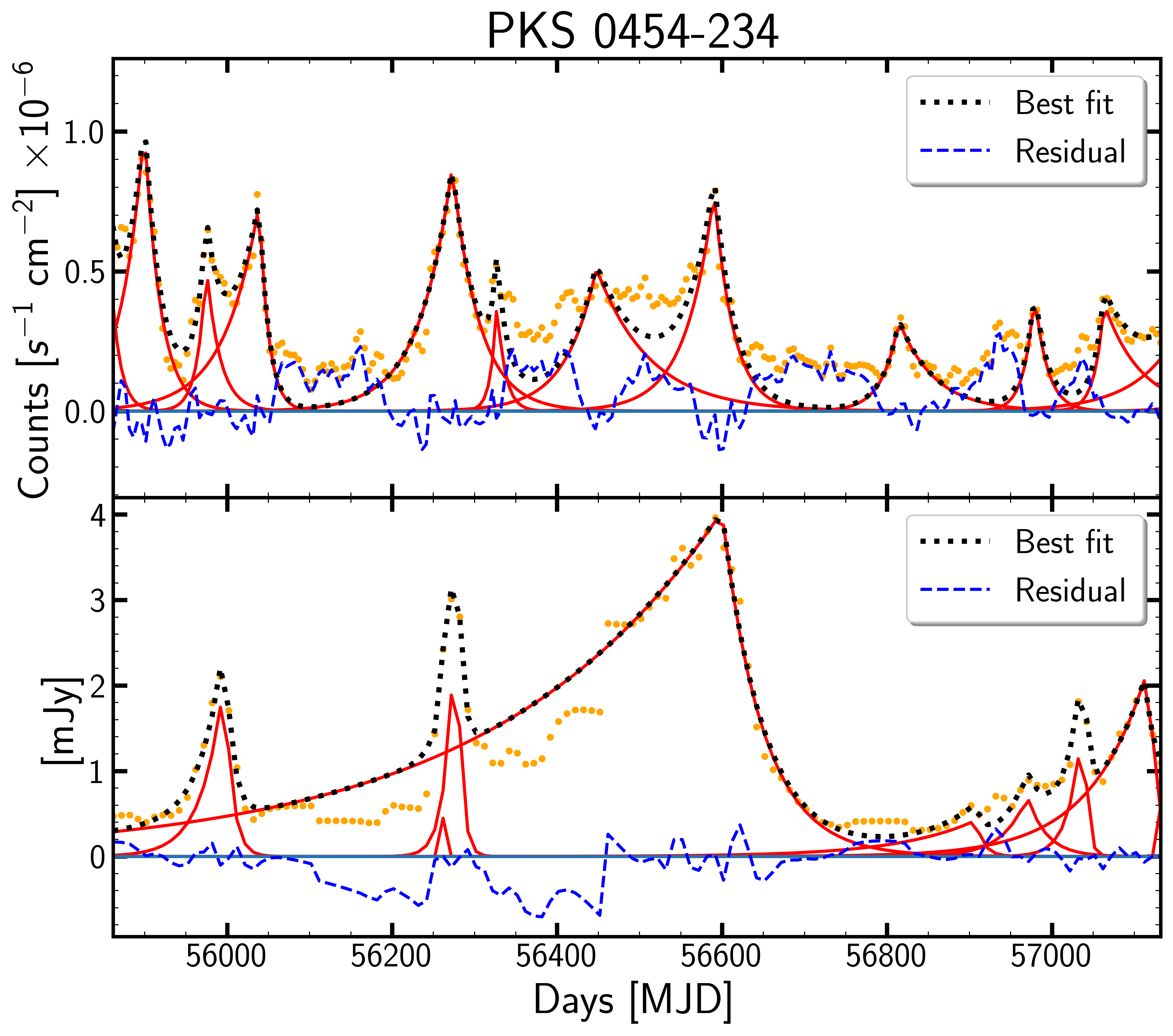}
        \includegraphics[scale = 0.23]{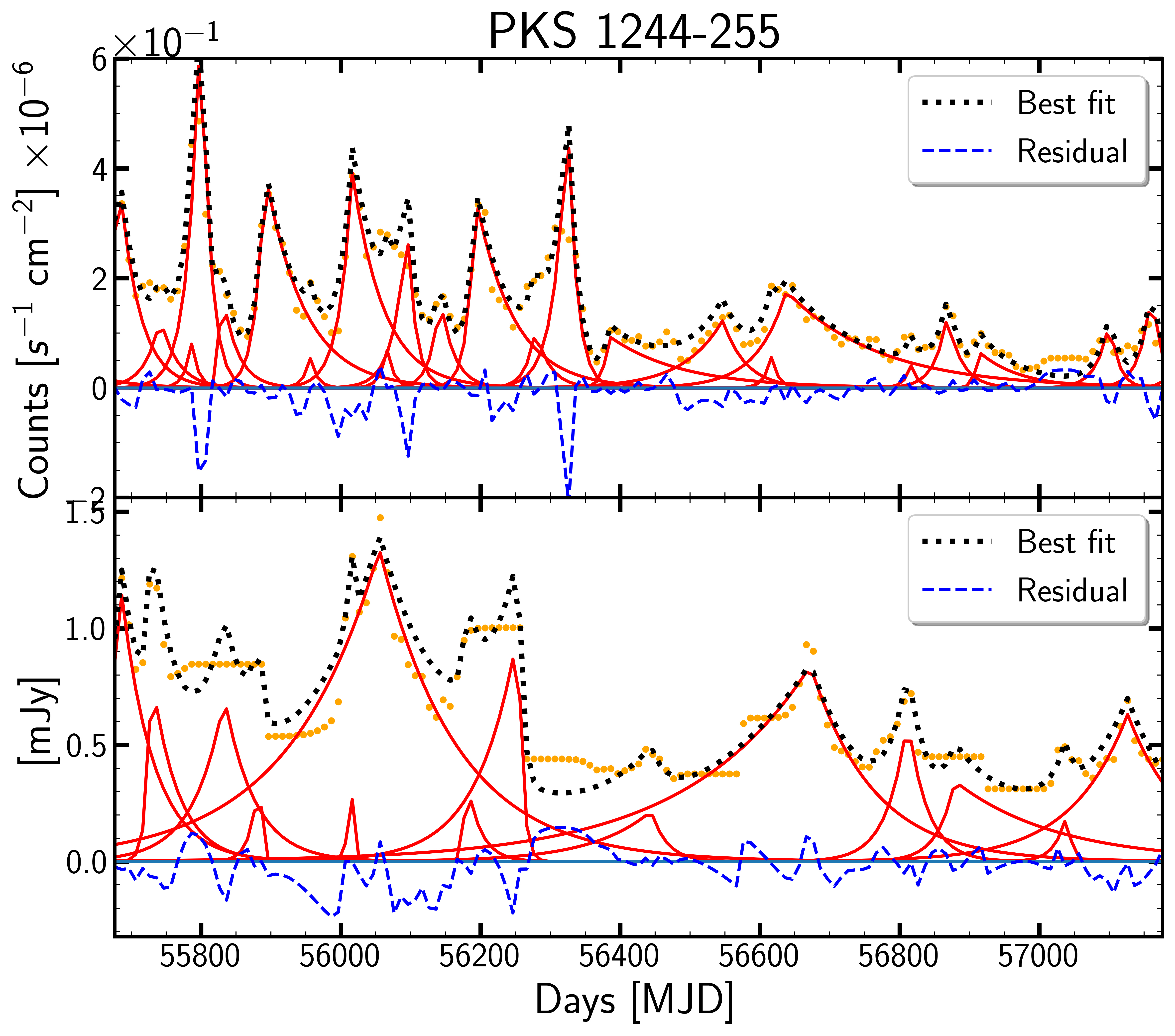}
        \includegraphics[scale = 0.23]{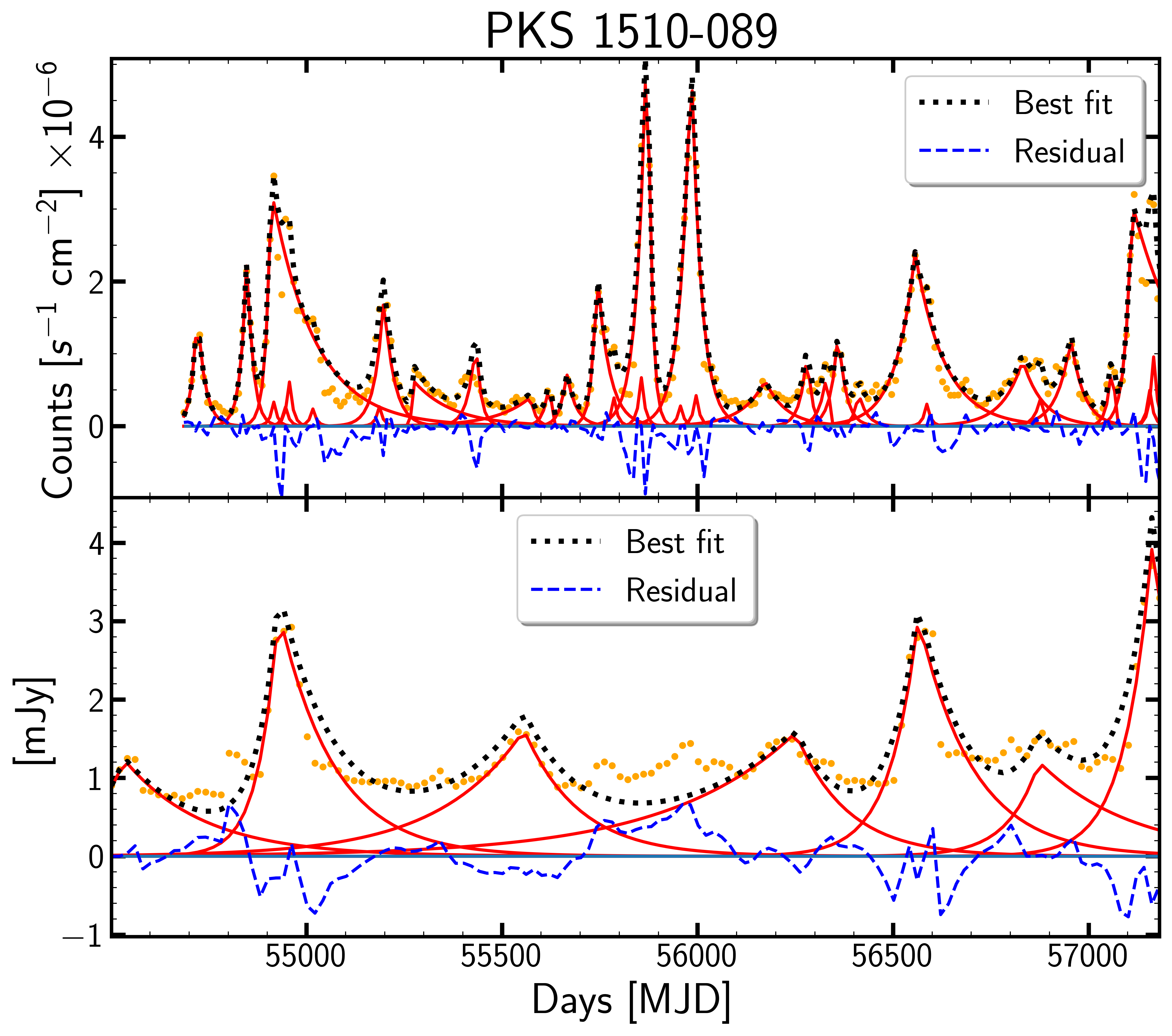}
        \includegraphics[scale = 0.23]{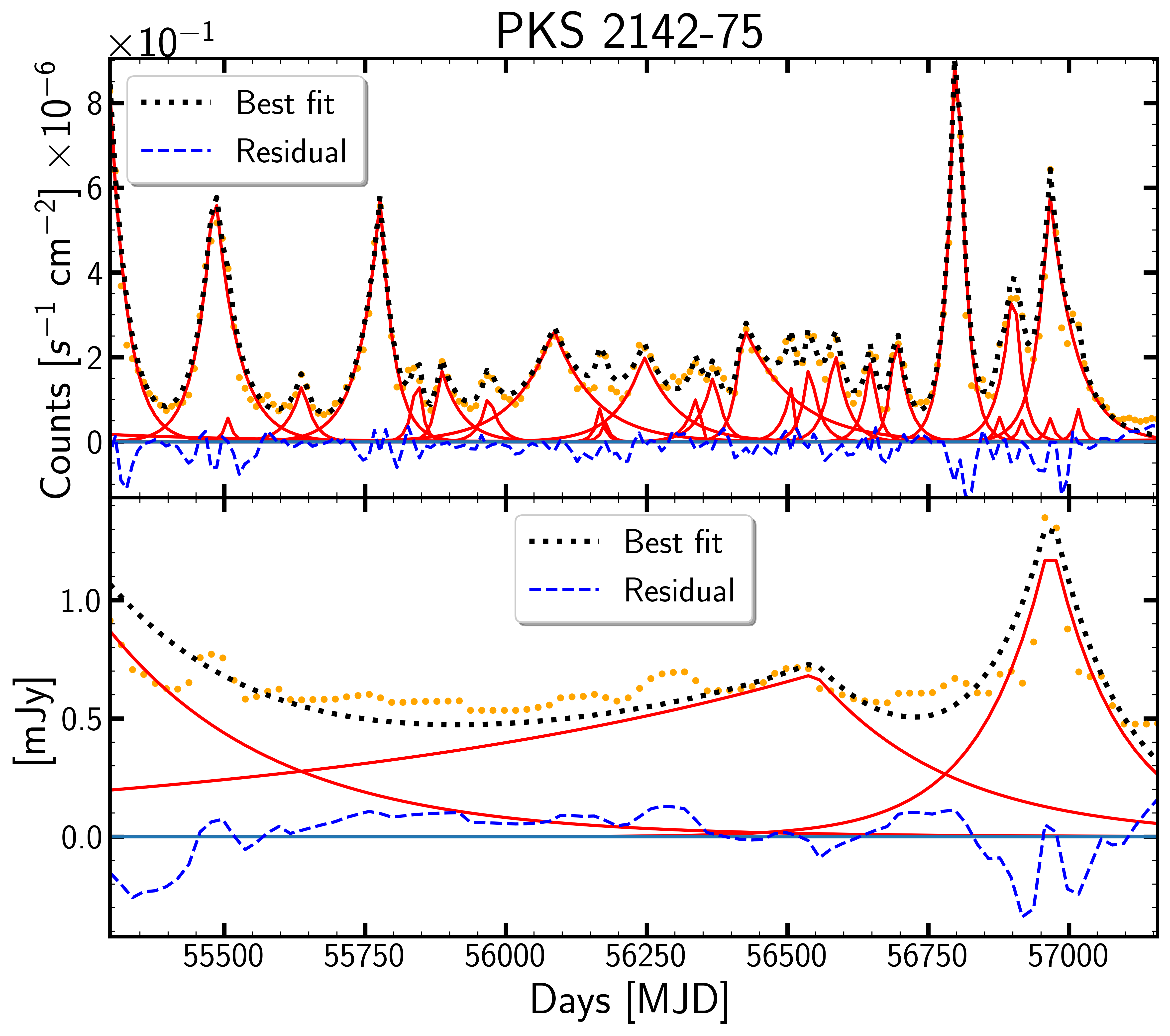}
        \includegraphics[scale = 0.23]{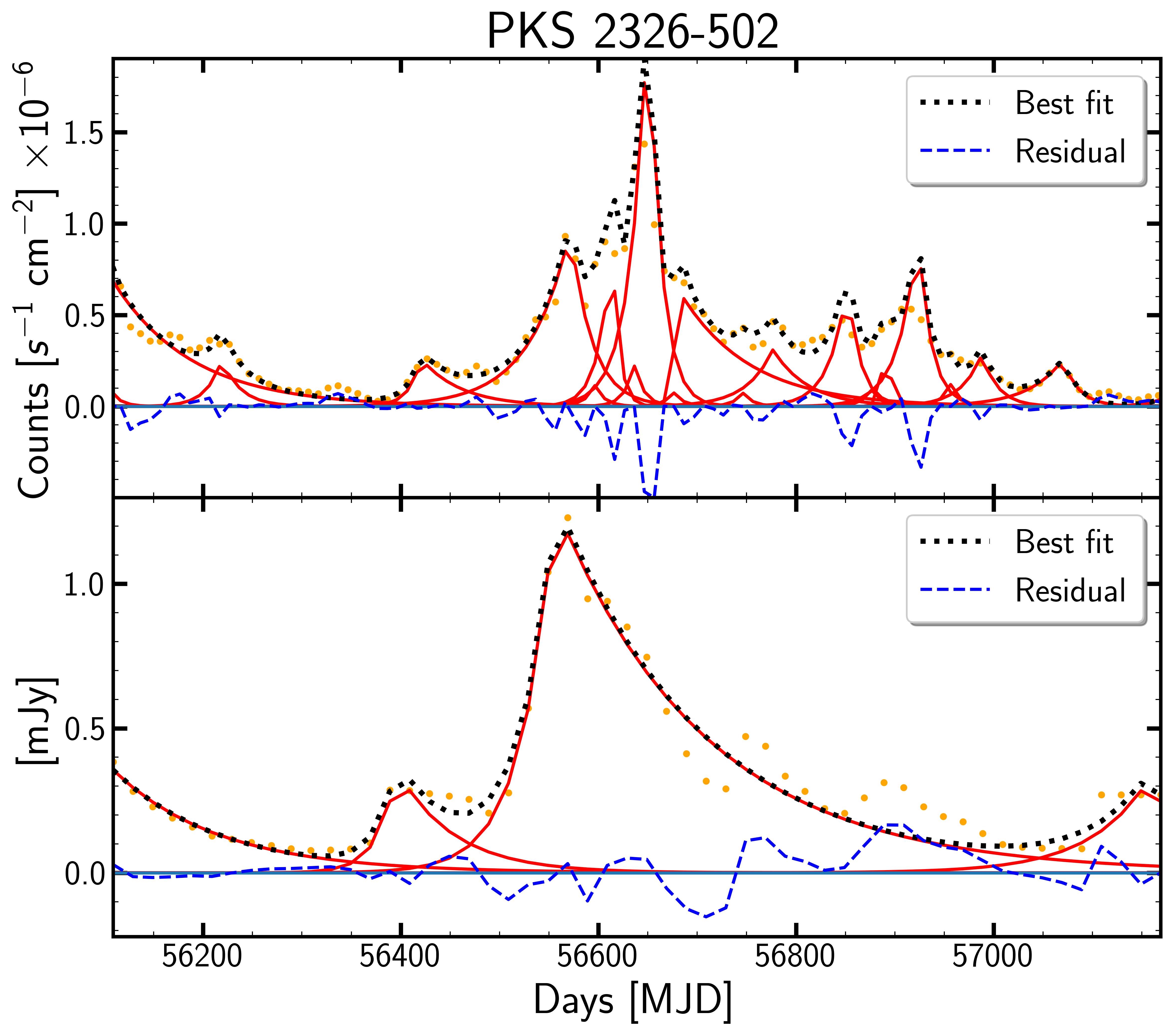}
        \includegraphics[scale = 0.23]{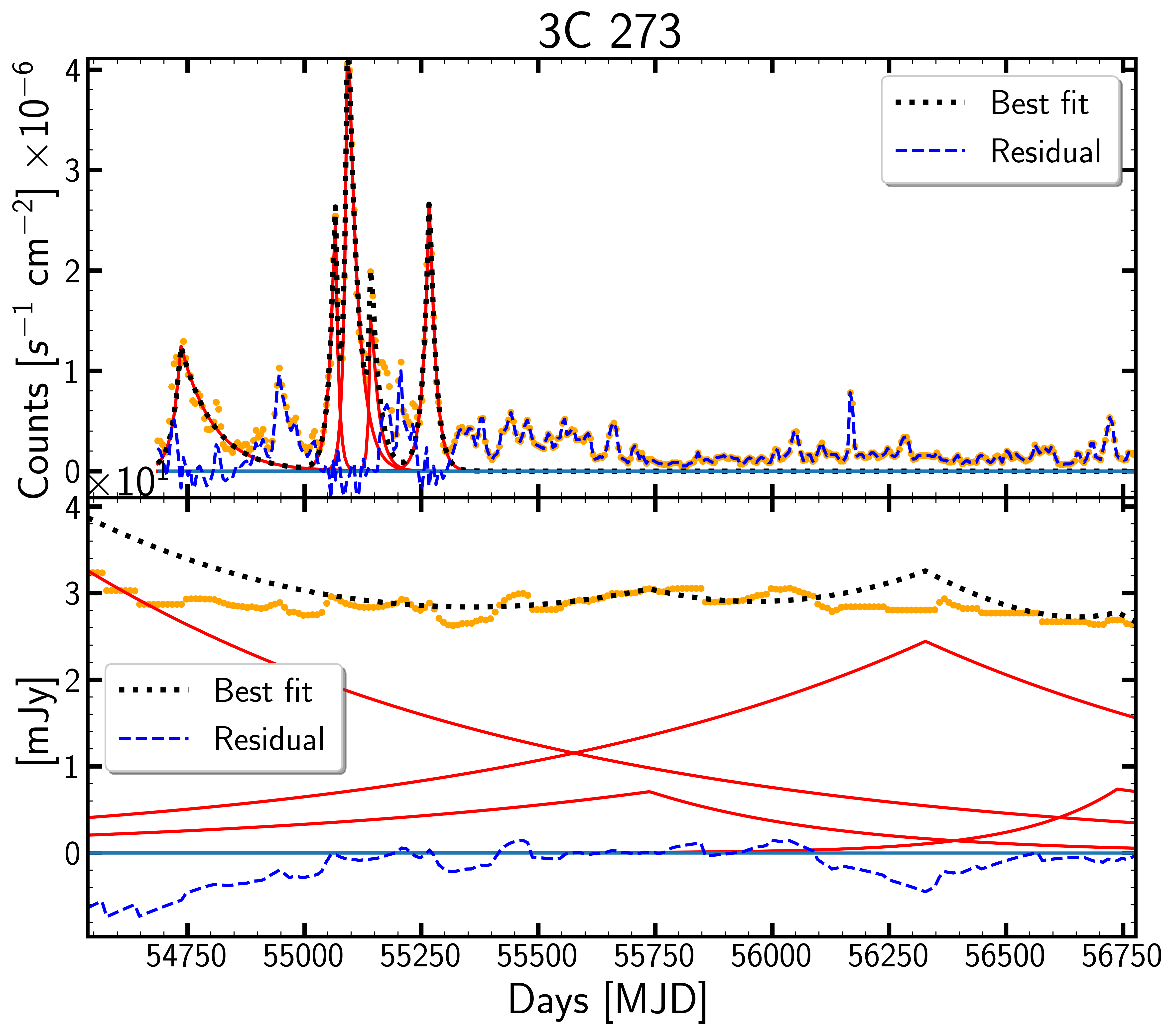}
        \includegraphics[scale = 0.23]{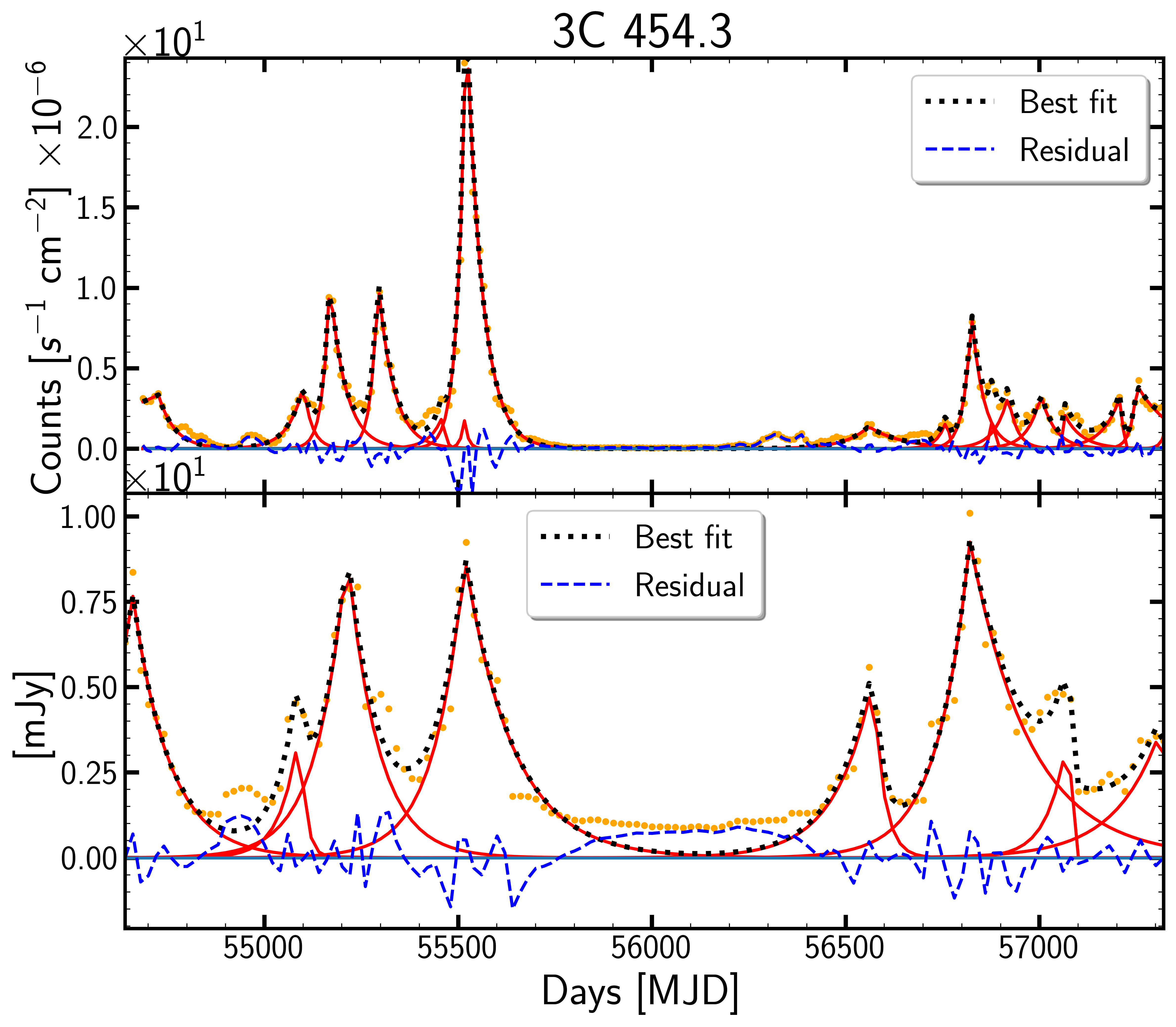}
    \end{center}
    \caption{`Decomposed' light curves, with individual flares shown as red lines, for the blazars in our sample. The name of the blazar is indicated in each plot. The top light curve in each plot is at the $\gamma$-ray band, and the bottom one is at the $R$ band. The black dotted line in each plot shows the best-fit sum of all flares. The source is indicated at the top of each plot.}
    \label{fig:fit_LCs}
\end{figure*}

In a smoothed light curve, we first identify the highest (most luminous) flare and fit it with Equation \ref{eqn:flare}. The fitted flare is subsequently subtracted from the total light curve. From the resultant light curve, we identify the new highest flare (thus, the second highest flare in the original light curve). We perform the fit again and subtract it. {This process finds flares with decreasing amplitudes in each step. It is repeated till the highest flare left in the residual light curve is $\le 10\%$ of the highest flare in the original light curve. Energy dissipation in flares with amplitude smaller than that may be neglected compared to the larger outbursts and will not have any significant effect in our final results.} {The results of this procedure on the sources in our sample are shown in Figure \ref{fig:fit_LCs}.}

The energy emitted by each individual flare is calculated by integration to find the area under the best-fit model describing the flare. We choose a 20-day window and select `flare pairs' to find the ratio of energy dissipated in a flare pair's optical and $\gamma$-ray components, \textit{i.e.}, the $t_0$ values, obtained by the best fit, of each member in the pair should not be more than 20 days apart. This ensures that the flares in both bands have been produced by the same event.


By the flare decomposition method, we have found a total of 47 flare pairs across 10 blazars. In order to determine the $\gamma$-ray to optical energy ratio, we convert the energy dissipated in the $\gamma$-ray flares from photon to energy units using the same method as described in \citet{Barat2022}. The values of the photon indices required for the calculation are obtained from the {4FGL-DR3} catalog \citep{Abdollahi2020_4FGL, Abdollahi2022}. The value of the $\gamma$-ray to optical energy ratio of the flare-pairs found in our analyses are between 0.01 and 100 with most values being in the range $0.1-1$. 

The uncertainty in the emitted energy of a flare has been estimated by integrating the flare function, $f(t)$ (Equation \ref{eqn:flare}) between $t_0 - \tau_{\rm rise} - \delta \tau_{\rm rise}$ to $t_0 + \tau_{\rm fall} + \delta \tau_{\rm fall}$. Here $\delta \tau_{\rm rise}$ and $\delta \tau_{\rm fall}$ are the uncertainties in the rising and decaying timescales, respectively, and are obtained from the covariance matrix of the fit. Then the integral from $t_0 - \tau_{\rm rise}$ to $t_0 + \tau_{\rm fall}$ has been subtracted from the above integral to find the resultant uncertainty.

\subsection{Numerical simulation of the blazar jet emission}


To determine the $\gamma$-ray emitting region from the energy ratios of the observed flare pairs, we perform a simulation of the emission from the relativistic jet. We consider in this simulation that the optical radiation is dominated by the synchrotron process and the $\gamma$-ray emission is dominated by the external Compton process with seed photons from the BLR and the DT, in accordance with our sample of sources which are all low synchrotron peaked (LSP) blazars, in which the contribution from the synchrotron self-Compton process falls off at around X-rays, i.e.,  at frequencies lower than that of the second hump in the SED. This numerical model has been described in \citet{Majumdar2019}, \citet{AritraKundu2022}, and \citet{Barat2022}. Below we describe it with more details, and with some new features. 

The jet is assumed to be cylindrical in structure, and is divided into multiple `cells', which are also cylindrical. Each cell has its own magnetic field ($B$), which scales as, $B \sim r^{-1}$, where $r$ is the distance from the central SMBH. The longer timescale emission from the jet is attributed to the passage of shocks along the jet \citep{Malzac2012, Joshi2014}. A shock passing through a cell energizes the electron-distribution to a power-law:
\begin{equation}
    N(\gamma) = N_0 \gamma^{-s}
\end{equation}

We simulate the SSC emission with a seed photon intensity derived from synchrotron radiation, with the emissivity:
\begin{equation}
    j_s(\nu) = \frac{\sqrt{3} e^2 B}{4 \pi m_e c^2} \int_{\gamma_{\rm min}}^{\gamma_{\rm max}} {\rm d}\gamma~N(\gamma)~F\left(\frac{\nu}{\nu_c}\right)
\end{equation}
where:
\begin{equation}
    F(x) = x~\int_x^{\infty} K_{5 / 3}(z)~{\rm d}z,
\end{equation}
$K_{5 / 3}(z)$ being the modified Bessel function of the second kind of order $\frac{5}{3}$, and $\nu_c$ is the critical frequency given by:
\begin{equation}
    \nu_c = \frac{2 B \gamma^2 e}{4 \pi m_e c}
\end{equation}
$F(x)$  is computed using a piecewise continuous polynomial fitting function that has maximum deviation of 0.00003 dex from the true value in the relevant range.

The inverse-Compton emissivity (in units of ${\rm erg} \cdot {\rm cm}^{-3} \cdot {\rm s}^{-1} \cdot {\rm Hz}^{-1} \cdot {\rm sr}^{-1}$), from the same electron-distribution as that generating the synchrotron emission, is calculated as:
\begin{equation}
    j_{\rm IC}(\nu) = \int_{\gamma_{\rm min}}^{\gamma_{\rm max}} {\rm d}\gamma~N(\gamma)~\int {\rm d}\nu_i~I_{\rm seed}(\nu_i) \sigma_{\rm IC}(\nu, \nu_i, \gamma),
\end{equation}
where $\sigma_{\rm IC}$ is the Compton cross-section given by:
\begin{equation}
    \sigma_{\rm IC}(\nu, \nu_i, \gamma) = \frac{3}{32} \cdot \frac{\sigma_T x}{\nu_i} \left\{8 + 2 x - x^2 + 4 x \ln\left(\frac{x}{4}\right)\right\},
\end{equation}
with $x = \nu / (\nu_i \gamma^2)$, and $\nu_i$ is the frequency of the seed photons.

The SSC contribution from a particular cell is calculated as the sum of the contribution from that cell (`1-cell term') and that from all other cells (`2-cell term'). For the $m^{\rm th}$ cell, the monochromatic photon energy density due to the 1-cell term is:
\begin{equation}
    U_1(\nu_s) = \frac{L(\nu_s)}{V} \langle t_{\rm esc} \rangle = \langle t_{\rm esc} \rangle \int j_s(\nu_s)~{\rm d}\Omega,
\end{equation}
which gives a monochromatic intensity of:
\begin{equation}
    I_1(\nu_s) = j_s(\nu_s) \langle t_{\rm esc} \rangle c
\end{equation}

For the other ($n \neq m$) cells, the contribution from each cell is calculated as:
\begin{equation}
    I_2 = \frac{j_s(\nu_s)~\lambda(R, \delta x, z)~\Omega(R, \delta x, z)}{4 \pi},
\end{equation}
where $\lambda(R, \delta x, z)$ is the mean free path of the photon through the $n^{\rm th}$ cell, and $\Omega(R, \delta x, z)$ is the solid angle subtended by the $n^{\rm th}$ cell at the position of the $m^{\rm th}$ cell. They are functions of the radius of the cells ($R$), width of the cells ($\delta x$), and the distance between the $m^{\rm th}$ cell and the $n^{\rm th}$ cell ($z$). Thus, the total SSC intensity is:
\begin{equation}\label{eqn:I_SSC}
    I_{\rm SSC}(\nu_s) = I_1(\nu_s) + \sum_n I_{2}^{(n)}(\nu_s)
\end{equation}
The sum in Equation \ref{eqn:I_SSC} also incorporates light travel time effects. For each spatial point $x_m$ and time $t_p$, the $n \neq m$ term is computed for all $\{x_n, t_q\}$ coordinates for which:
\begin{equation}
\label{eqn:causality}
    |x_m - x_n| = c~(t_p - t_q)
\end{equation}

\begin{figure}
    \centering
    \includegraphics[width = \columnwidth]{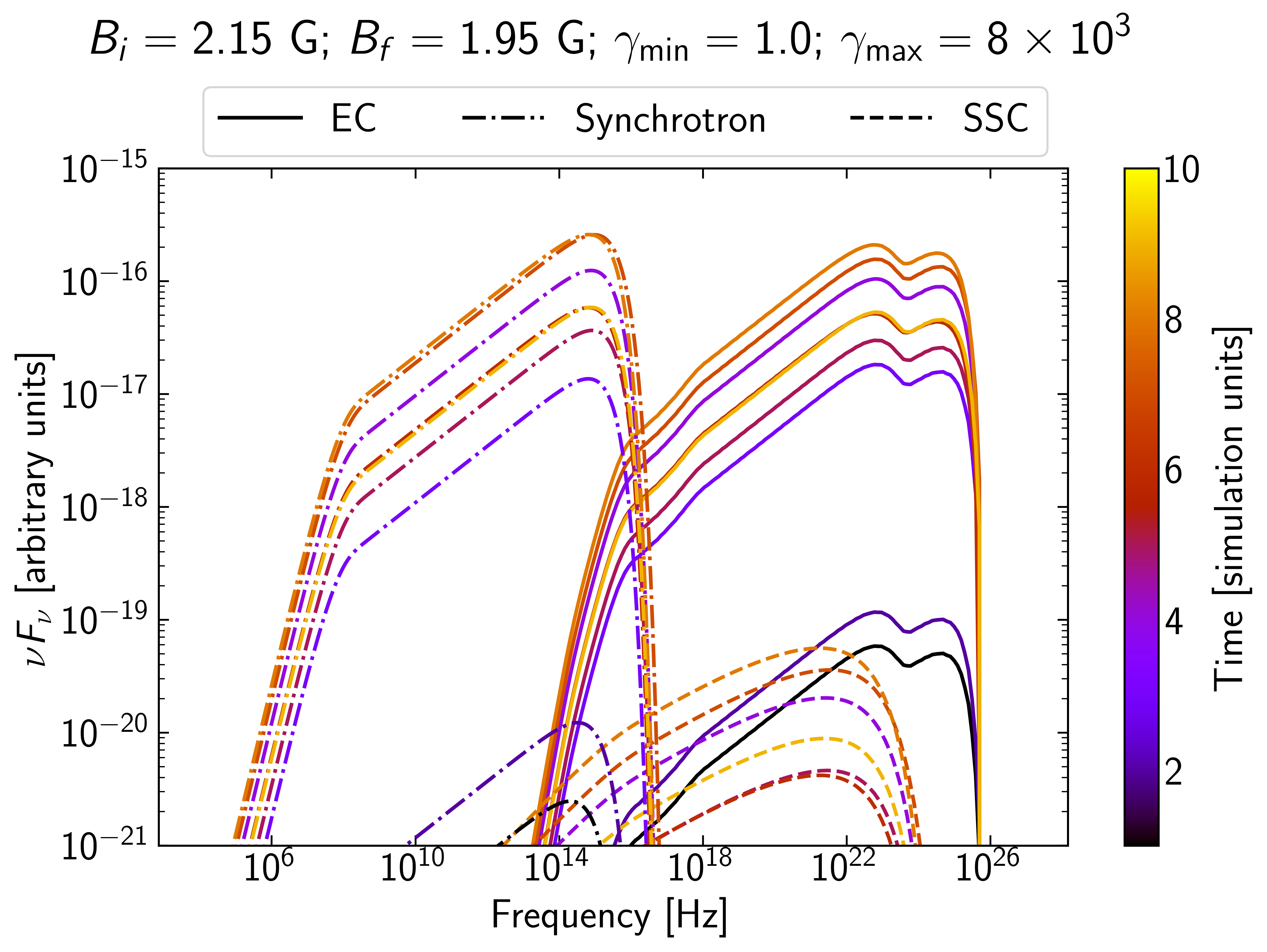}
    \caption{{The time-evolution of the simulated SEDs. For each case, the standard double humped structure is reproduced. Each component of the emission is plotted separately. The smaller bumps on the major peaks are due to the superposition of emissions from different cells in the jet. This figure depicts simulated SEDs equally spaced in time for the entire duration of the simulation.}}
    \label{fig:simSweep}
\end{figure}

\begin{figure}
    \centering
    \includegraphics[width = \columnwidth]{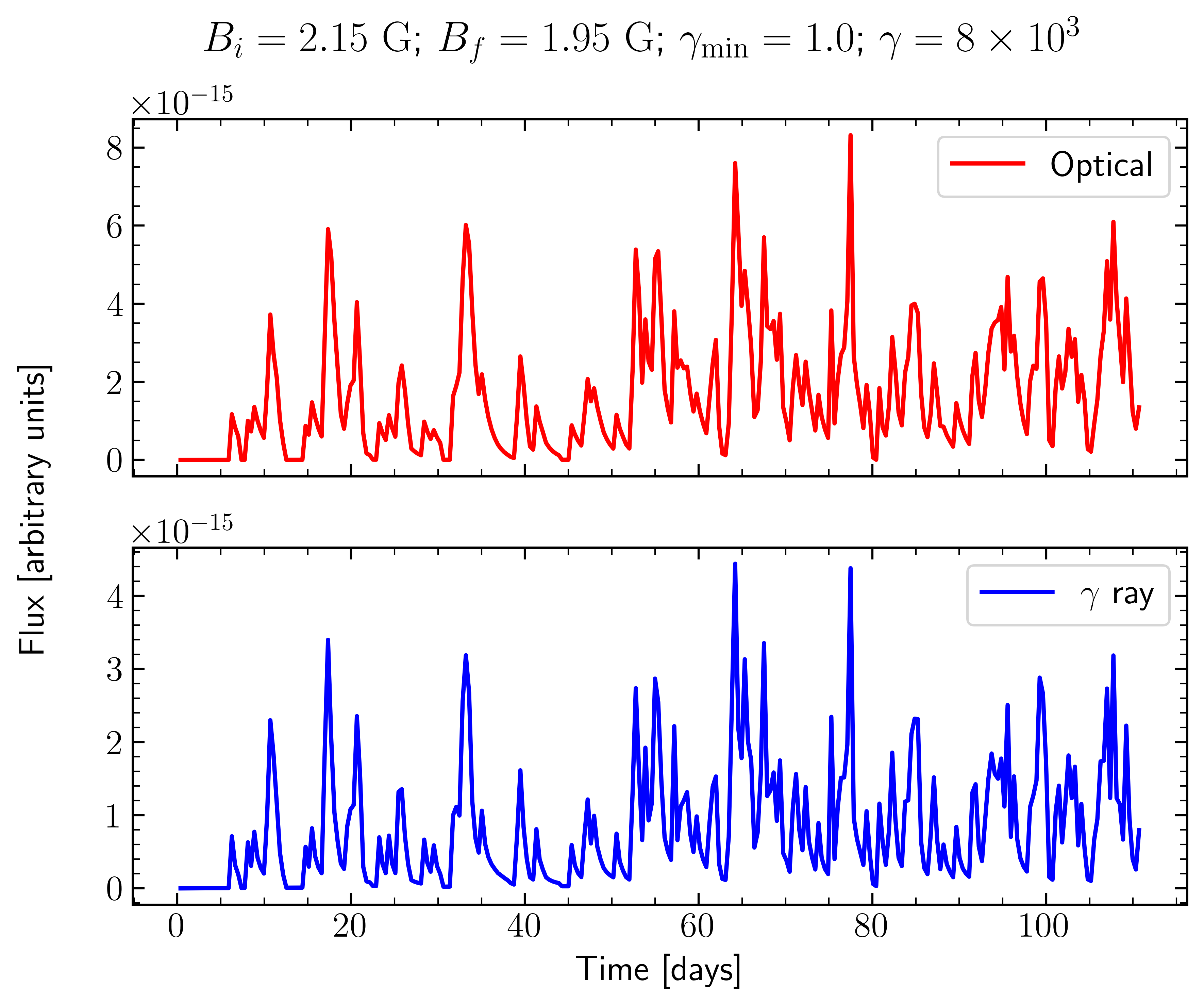}
    \caption{Months-timescale Light Curves simulated with the Jet Simulation: Optical \textit{(top)} and $\gamma$-ray \textit{(bottom)}. The parameters for the simulation are indicated at the top. The correlation between different bands can also be seen.}
    \label{fig:simulated_LC}
\end{figure}

The seed photon field intensity --- which drives the external Compton process --- is calculated as:
\begin{equation}\label{eqn:u_EC_BLR}
    U_{\rm BLR}(r) = \frac{\mathcal{E}_{\rm BLR} \Gamma_{\rm jet}^2 L_D}{3 \pi R_{\rm BLR}^2 c \left[1 + (r / R_{\rm BLR})\right]^{\beta_{\rm BLR}}}
\end{equation}
and
\begin{equation}\label{eqn:u_EC_DT}
    U_{\rm DT}(r) = \frac{\mathcal{E}_{\rm DT} \Gamma_{\rm jet}^2 L_D}{3 \pi R_{\rm DT}^2 c \left[1 + (r / R_{\rm DT})\right]^{\beta_{\rm DT}}},
\end{equation}
following \citet{Hayashida2012}. Here, $\mathcal{E}$ represents the fraction of the accretion disk luminosity (denoted by $L_D$) reprocessed by the BLR or the DT, and $\beta = v / c$. $R_{\rm BLR}$, $R_{\rm DT}$, and $r$ denote the distance of the BLR, DT, and the emission region, respectively, from the central SMBH. $\Gamma_{\rm jet}$ is the bulk Lorentz factor of the jet. We focus only on the seed photon populations from the BLR and the DT as that from the accretion disk is debeamed due to the emitting region moving along the jet away from the disk, and thus the accretion disk contribution is subdominant compared to the BLR or DT contribution. For the simulation, we have fixed the values of the magnetic field at the base ($B_i$) and the tip ($B_f$) of the emitting region, and assume the magnetic field to be smoothly varying between these two points. {Figure \ref{fig:simSweep} shows broadband SEDs generated using our simulation, where we show SEDs equispaced in time for one run of the simulation to illustrate the time evolution of the SEDs. We can see that the standard `double humped' nature of blazar SEDs is correctly reproduced using our simulation.}
Figure \ref{fig:simulated_LC} shows two light curves --- one in $R$ band and one in $\gamma$-ray band --- generated using our simulation.

\subsubsection{Turbulence in the Jet Magnetic Field}


In addition to a smoothly varying ($\sim 1 / r$) magnetic field, we implement turbulence by adding fluctuations in the magnetic field. The emission region is divided into $N$ number of cells. We assume that the fluctuations in the magnitude and direction of the magnetic field in neighboring cells may be partially correlated.
Thus, the number of such correlated zones is:
\begin{equation}
    N_{\rm zone} = \frac{x_{\rm max}}{L_{\rm corr}},
\end{equation}
where $x_{\rm max}$ is the maximum number of cells in the emission region and $L_{\rm corr}$ is the length of each zone in cell units.

The magnetic field at the $k^{\rm th}$ zone is given by:
\begin{equation}
    B_k = B_i + \frac{k}{N_{\rm zone}}~(B_f - B_i), \hspace{0.2 in} k = 1, 2, ..., N_{\rm zone}.
\end{equation}
We add fluctuations to the magnetic field in the $k^{\rm th}$ correlated zone as follows:
\begin{equation}
    B_k^{\prime} = B_k~[1 + 2~\Sigma~\{0.5 - {\mathcal{U}}(0, 1)\}],
\end{equation}
where $\Sigma$ is a parameter to normalize the amplitude of fluctuations in each zone, and we draw a random number between 0 and 1 from a uniform distribution ${\mathcal{U}}(0, 1)$.

Assuming an initial orientation $\theta_0$ of the magnetic field, we give a twist $\theta_t$ to this orientation at each cell as follows:
\begin{equation}
    \theta_t = \{0.5 - {\mathcal{U}}(0, 1)\} \cdot 2~\theta_{\rm max},
\end{equation}
where $\theta_{\rm max}$ is the maximum angle of twist. $\theta_t$ represents a rotation of the angle of the magnetic field at each cell, and is updated at each correlated zone. Thus, the angle of the magnetic field at the $j^{\rm th}$ cell in a zone is:
\begin{equation}
    \theta_j = \theta_0 + j \theta_t, \hspace{0.5 in} j = 1, 2, ..., L_{\rm corr}.
\end{equation}
Before proceeding to the next zone and updating the rate of rotation, we set $\theta_0$ to be equal to the angle of orientation of the magnetic field in the last cell of the previous zone. The magnetic field at the $n^{\rm th}$ cell is calculated as:
\begin{equation}
    B_n = B_k^{\prime}~[1 + 2 \sigma~\{0.5 - {\mathcal{U}}(0, 1)\}],
\end{equation}
where $\sigma$ is a parameter that is used to normalize the amplitude of fluctuations in each cell. {The values of $L_{\rm corr}$, $\sigma$, $\Sigma$ and $\theta_{\rm max}$ are decided based on qualitative comparison with observed light curves. A paper led by one of the coauthors is in preparation, in which a more detailed description of the model, including the implementation of the small-scale fluctuations, and overall comparison with multi-wavelength light curves will be described. }

\begin{table}
\begin{center}
\begin{tabular}{c|c}
\hline
\hline
        Parameter & Value \\
        \hline
        \hline
        $\Gamma_{\rm jet}$ & 10 \\
        $\gamma_{\rm min}$ & 1 \\
        $\gamma_{\rm max}$ & $8 \times 10^3$ \\
        Power law index, $s$ & 2.5 \\
        $B_i$ & 2.15 G \\
        $B_f$ & 1.95 G \\
        $L_{\rm D}$ & $10^{45}$ erg $\cdot$ s$^{-1}$ \\
        $\beta_{\rm BLR}$ & 3 \\
        $\beta_{\rm DT}$ & 4 \\
        $\theta_v$ & $5^{\circ}$ \\
        \hline
\end{tabular}
\end{center}
\caption{{Parameters used for generating the light curves of Figure \ref{fig:simulated_LC} using our simulation. $\gamma_{\rm min}$ and $\gamma_{\rm max}$ denote the lower and upper bounds of the electron energy distribution, respectively, and $\theta_v$ is the viewing angle to the blazar. All other parameters are as defined previously.}}
\label{tab:sim_params}
\end{table}



\section{GeV Emission Regions of the Blazars in Our Sample}
\label{sec:results}
    In this section, we apply the techniques discussed previously to the blazars in our sample to determine the locations of their $\gamma$-ray emitting regions, relative to the position of their BLR and torus. Certain parameters, which are kept constant in the simulations of light curves, are given in Table \ref{tab:const_prms}, with the corresponding values.

\begin{table}
    \centering
    \begin{tabular}{|c|c|}
    \hline
    \hline
       Parameter & Value \\
        \hline
        \hline
        Power-law index, $s$ & 2.5 \\
        $\beta_{\rm BLR}$ & 3 \\
        $\beta_{\rm torus}$ & 4 \\
        $\mathcal{E}_{\rm BLR}$ & 0.1 \\
        $\mathcal{E}_{\rm torus}$ & 0.01 \\
        \hline
    \end{tabular}
    \caption{The parameters, values of which are kept fixed in our simulation, with meanings as explained in the text.}
    \label{tab:const_prms}
\end{table}

    For each blazar, we follow a general procedure for extracting an estimate of the location of the $\gamma$-ray emission zone. First we extract all possible flare-pairs from the $R$-band and GeV light curves of the source. Then, we estimate the ratio of the energy emitted at $\gamma$-rays to that at the optical band, in each such pair. Using values of parameters of the jets obtained from the {literature \citep[\textit{e.g.},][]{Paliya_2017, Ghisellini2010, Dutka2017, Castignani2017}}, we simulate long term ($\sim$ 150--200 days) light curves of the relevant blazar. {Most of the parameter values we use are taken from \citet{Paliya_2017}, who derive those from modeling the SEDs of a large sample of \textit{Fermi} blazars in the CGRaBS Catalog \citep{CGRaBS}. We use those values as their sample is drawn from the latest 4FGL-DR3 data, which we also use and majority of the sources in our sample are included in their work.} 
    By the same method of flare decomposition, we extract the flares from the simulated light curves and obtain the $\gamma$-to-optical energy dissipation ratio averaged over all flare-pairs during the simulated interval of the relevant blazar. We repeat the same for several different trials, where at each trial we change the distance of the emission zone from the central engine. Finally, we see at what distances the simulated ratio matches with each observed ratio. We designate those distances as the originating location for that flare.

\begin{table*}
    \centering
    \begin{tabular}{|c|c|c|c|c|c|c|}
        \hline
        \hline
        Blazar Name & $\Gamma$ & $\gamma_{\rm min}$ & $\gamma_{\rm max}$ & $B$ [G] & $L_D$ [$10^{46}$ erg $\cdot$ s$^{-1}$] & $\theta_v$ [$^{\circ}$] \\
        \hline
        \hline
        PKS 0208-512 & {11.0} & 1.0 & $8 \times 10^3$ & {2.0} & {0.39} & 3.000 \\
        PKS 0402-362\footnote{\citet{Das2023}} & 16.2 & 49.1 & $10.2 \times 10^3$ & 2.50 & 5.00 & 0.059 \\
        PKS 0426-380 & {11.0} & 1.0 & $6 \times 10^3$ & {2.50} & f{1.00} & 3.000 \\
        PKS 0454-234 & f{13.0} & 1.0 & $4 \times 10^3$ & f{3.50} & {0.20} & 3.000 \\
        PKS 1244-255 & {11.0} & 1.0 & $4 \times 10^3$ & {3.30} & {0.79} & 3.000 \\
        PKS 1510-089 & 14.1 & 1.0 & $4 \times 10^3$ & 3.70 & 0.42 & 3.000 \\
        PKS 2142-75\footnote{\citet{Dutka2013_PKS-2142}} & 30.0 & 11.0 & $7 \times 10^3$ & 1.00 & 2.90 & 0.033 \\
        3C 454.3 & {18.0} & 1.0 & $4 \times 10^3$ & {3.00} & {1.20} & 3.000 \\
        PKS 2326-502 & {9.0} & 1.0 & $4 \times 10^3$ & {0.8} & {0.21} & 3.000 \\
        3C 273 & 12.9 & 1.0 & 2 $\times 10^4$ & 11.6 & 4.80 & 3.000 \\
        \hline
    \end{tabular}
    \caption{Physical parameters for our sample of blazars. These parameters have been obtained from {\citet{Paliya_2017}}, unless indicated otherwise. {The parameters for PKS 1510-089 and PKS 2142-75 have been taken from \citet{Ghisellini2010}.}}
    \label{tab:blazar_parameters}
\end{table*}

We set the location of the BLR and torus following scaling relations obtained from reverberation mapping analysis \citep[\textit{e.g.,}][]{Bentz2013}:
\begin{eqnarray}
    R_{\rm BLR} &=& 0.1~(L_{D, 46})^{0.5}~{\rm pc} \label{eqn:R_BLR} \\
    R_{\rm torus} &=& 2.5~(L_{D, 46})^{0.5}~{\rm pc} \label{eqn:R_torus}
\end{eqnarray}
where $L_{D, 46}$ denotes the accretion disk luminosity in units of $10^{46}~{\rm erg}\cdot{\rm s}^{-1}$.


\begin{figure*}
    \centering

        \includegraphics[scale = 0.4]{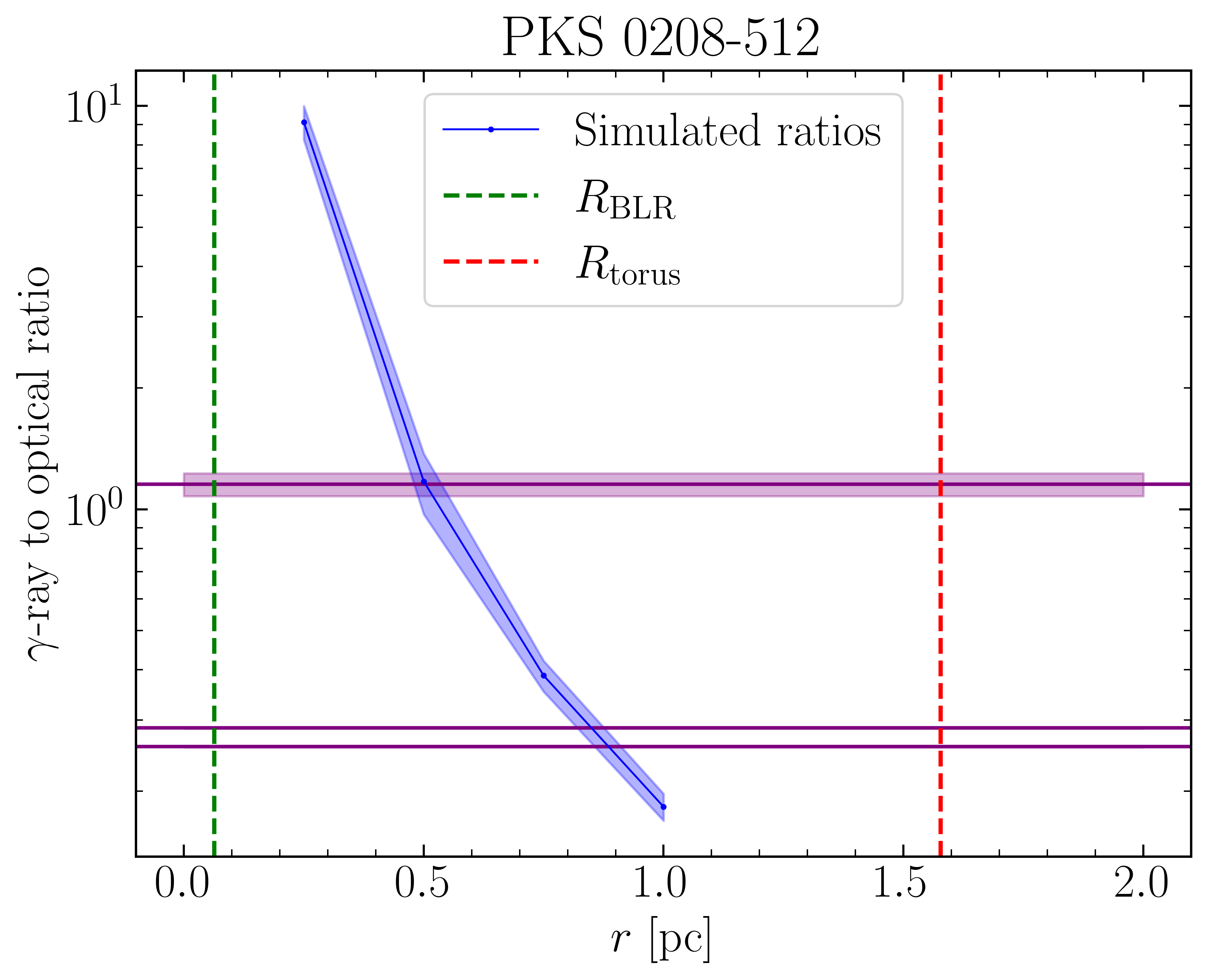}
        \includegraphics[scale = 0.4]{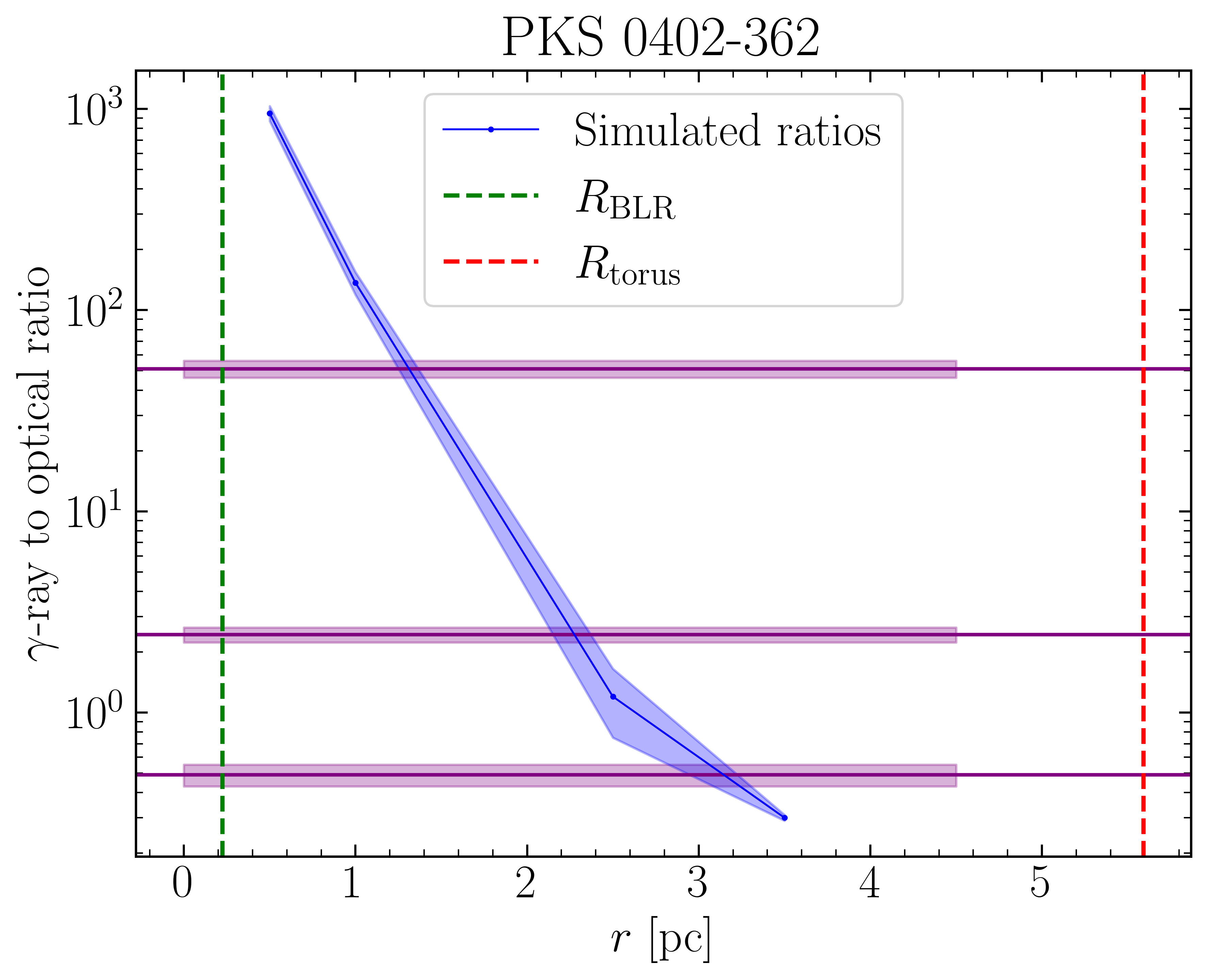}
        \includegraphics[scale = 0.4]{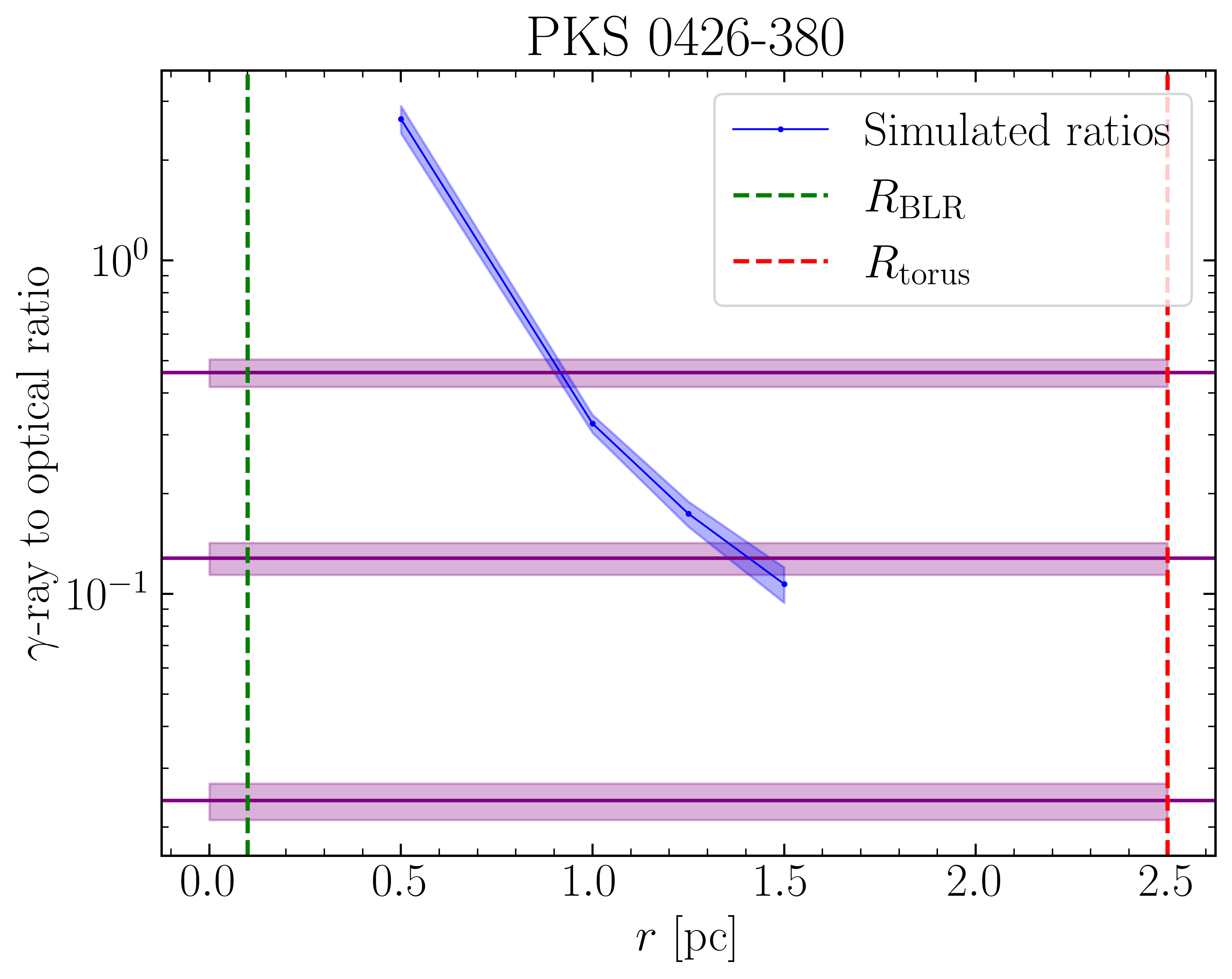}
        \includegraphics[scale = 0.4]{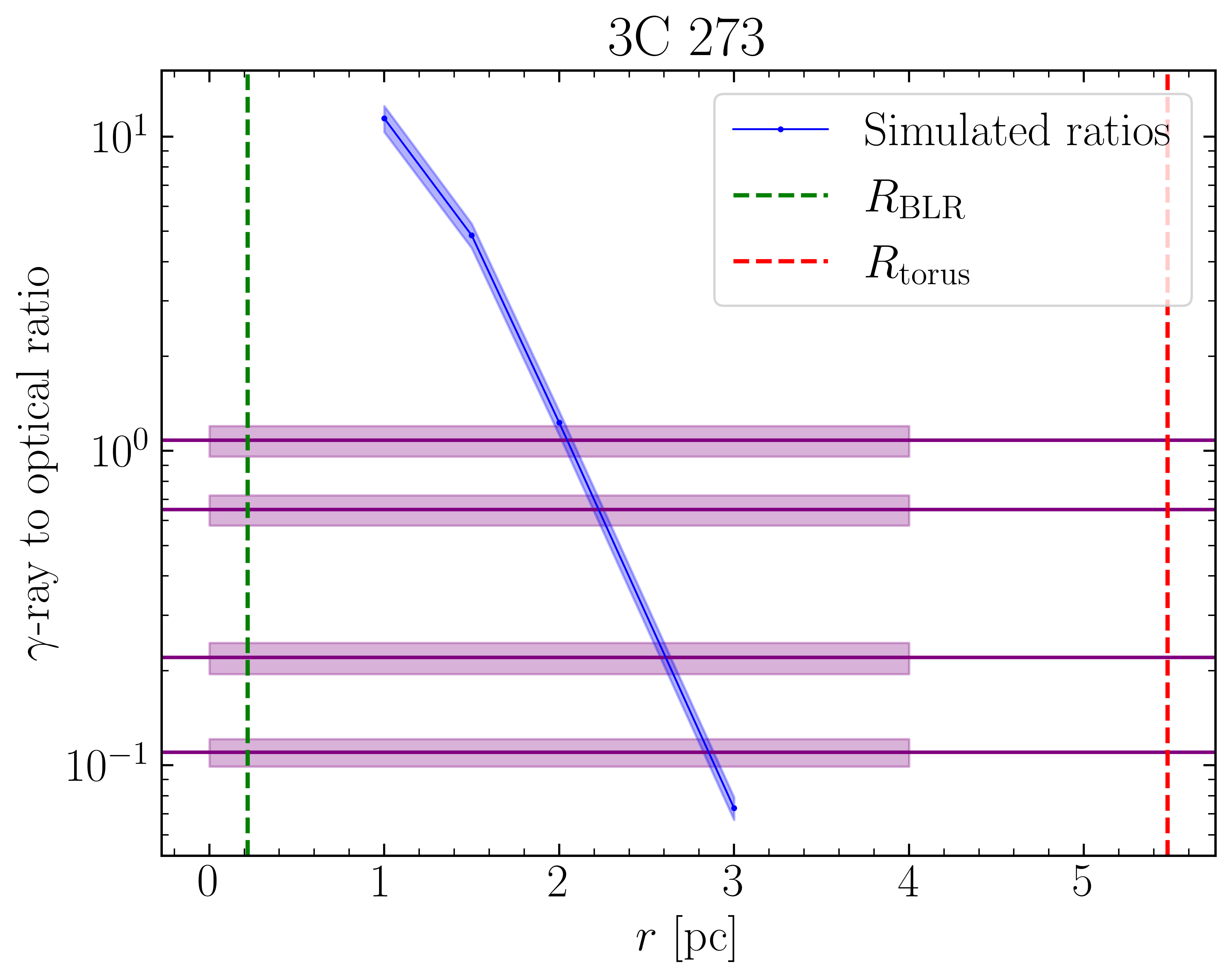}
        \includegraphics[scale = 0.4]{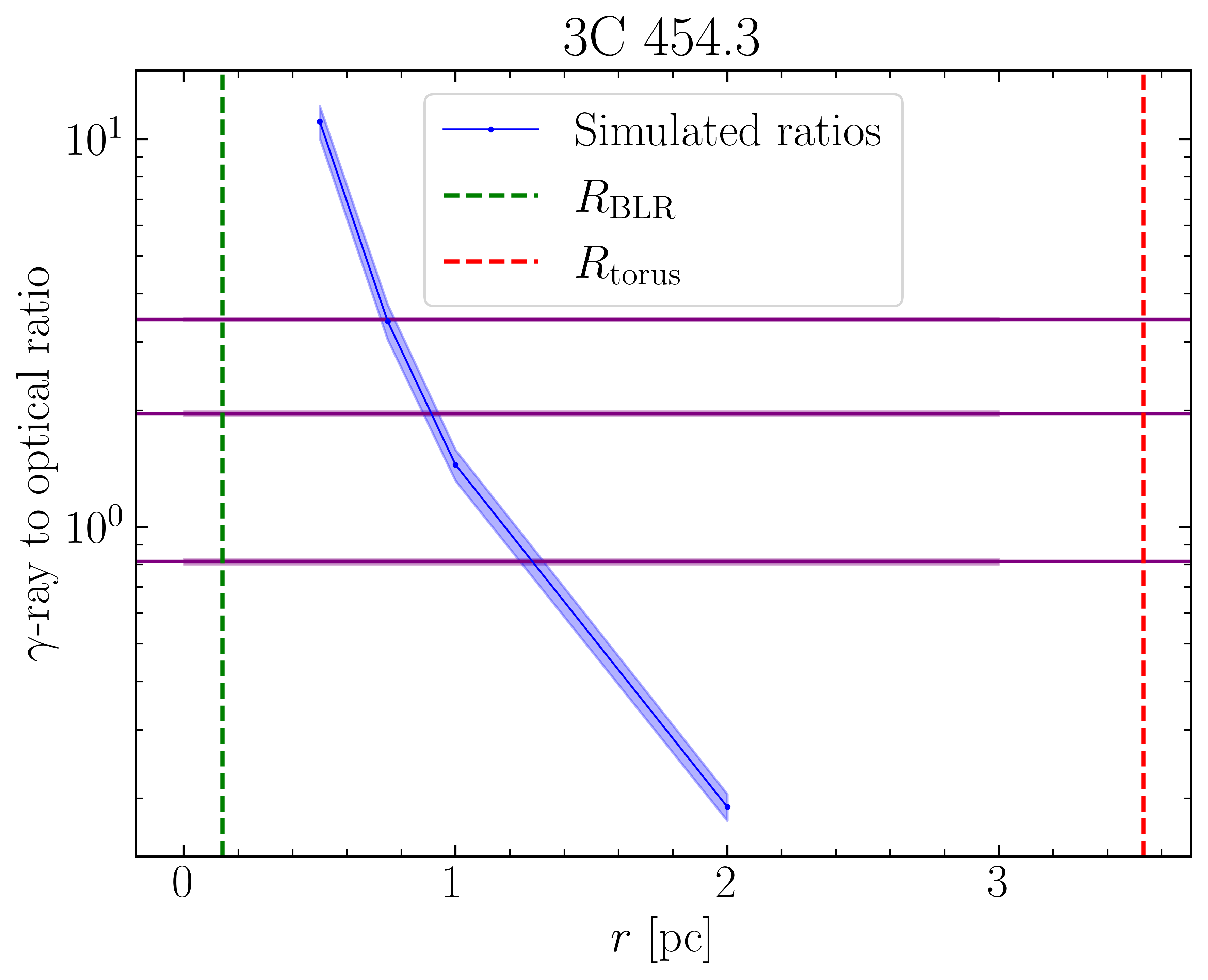}
    \caption{Variation of the simulated $\gamma$-to-optical ratio, as a function of the distance of the $\gamma$-ray emitting region from the central engine ($r$), is shown as the black dotted lines. The vertical green and red lines denote the position of the BLR and the DT, respectively. The horizontal lines in each plot denote the $\gamma$-to-optical ratio for each flare-pair of the relevant blazar. {The shaded areas represent the uncertainties for the observed and simulated flare-pairs.} The name of the blazar is indicated in the title of each plot.}
    \label{fig:ratio_plots}
\end{figure*}

\begin{figure*}
    \ContinuedFloat
        \includegraphics[scale = 0.4]{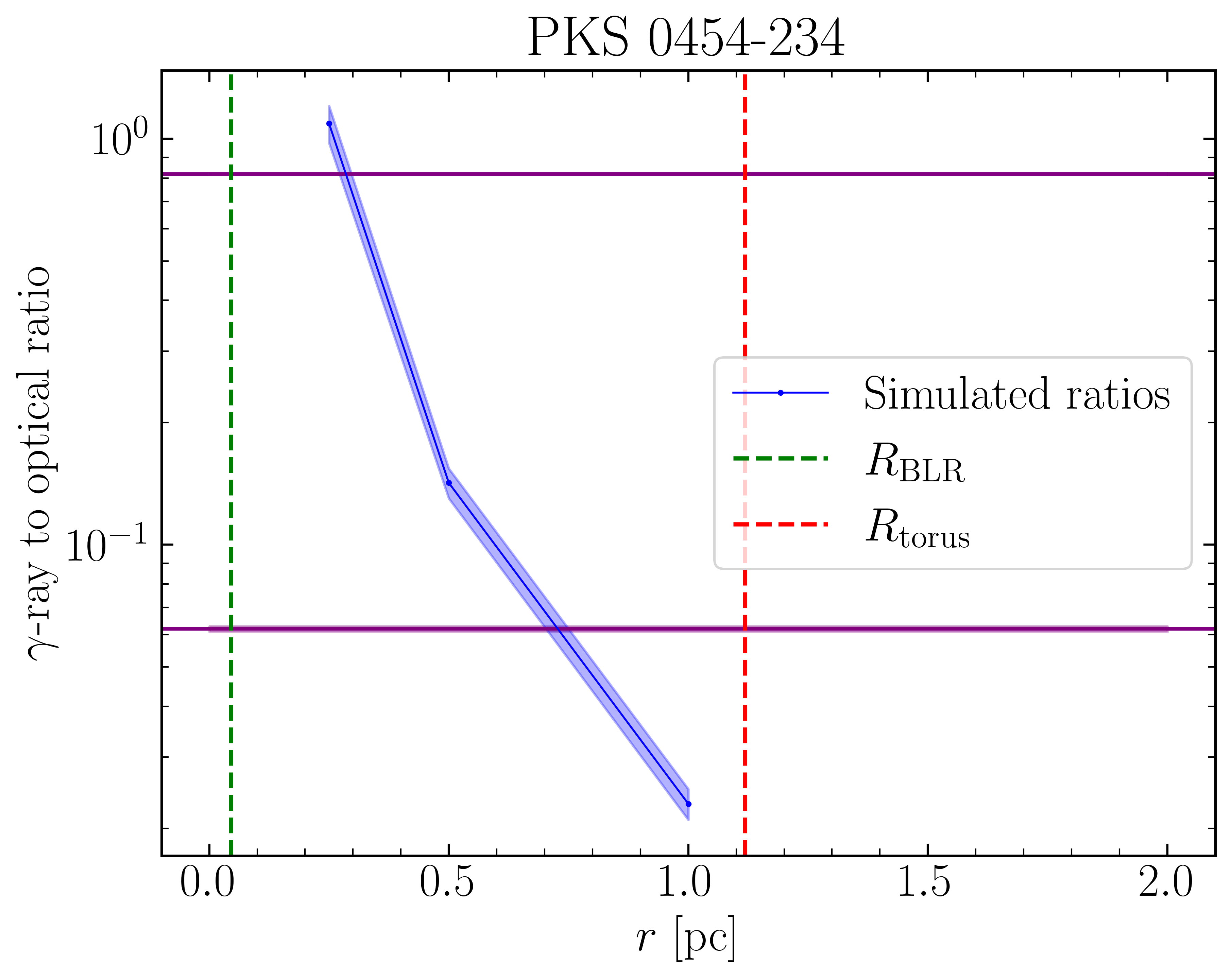}
        \includegraphics[scale = 0.4]{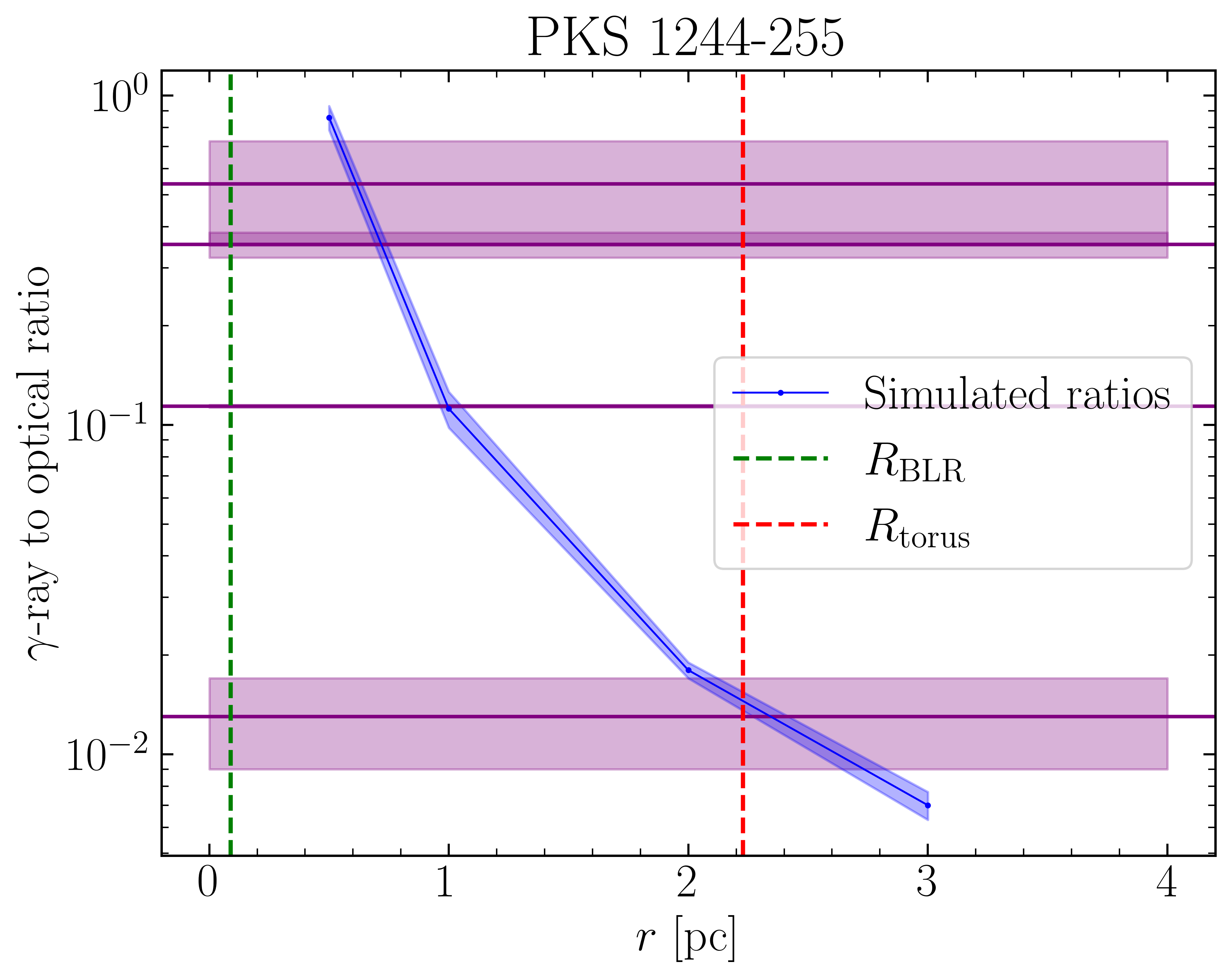}
        \includegraphics[scale = 0.4]{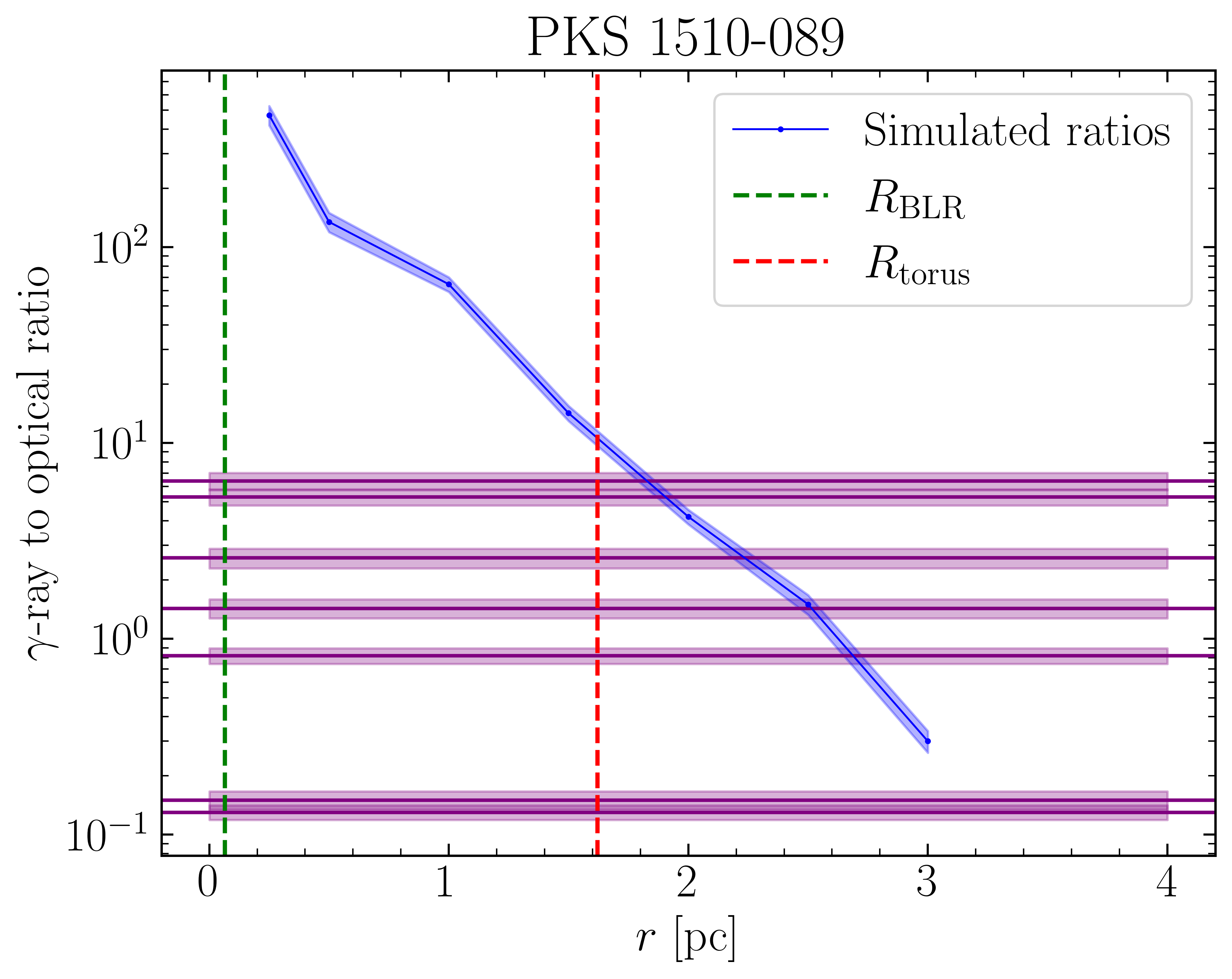}
        \includegraphics[scale = 0.4]{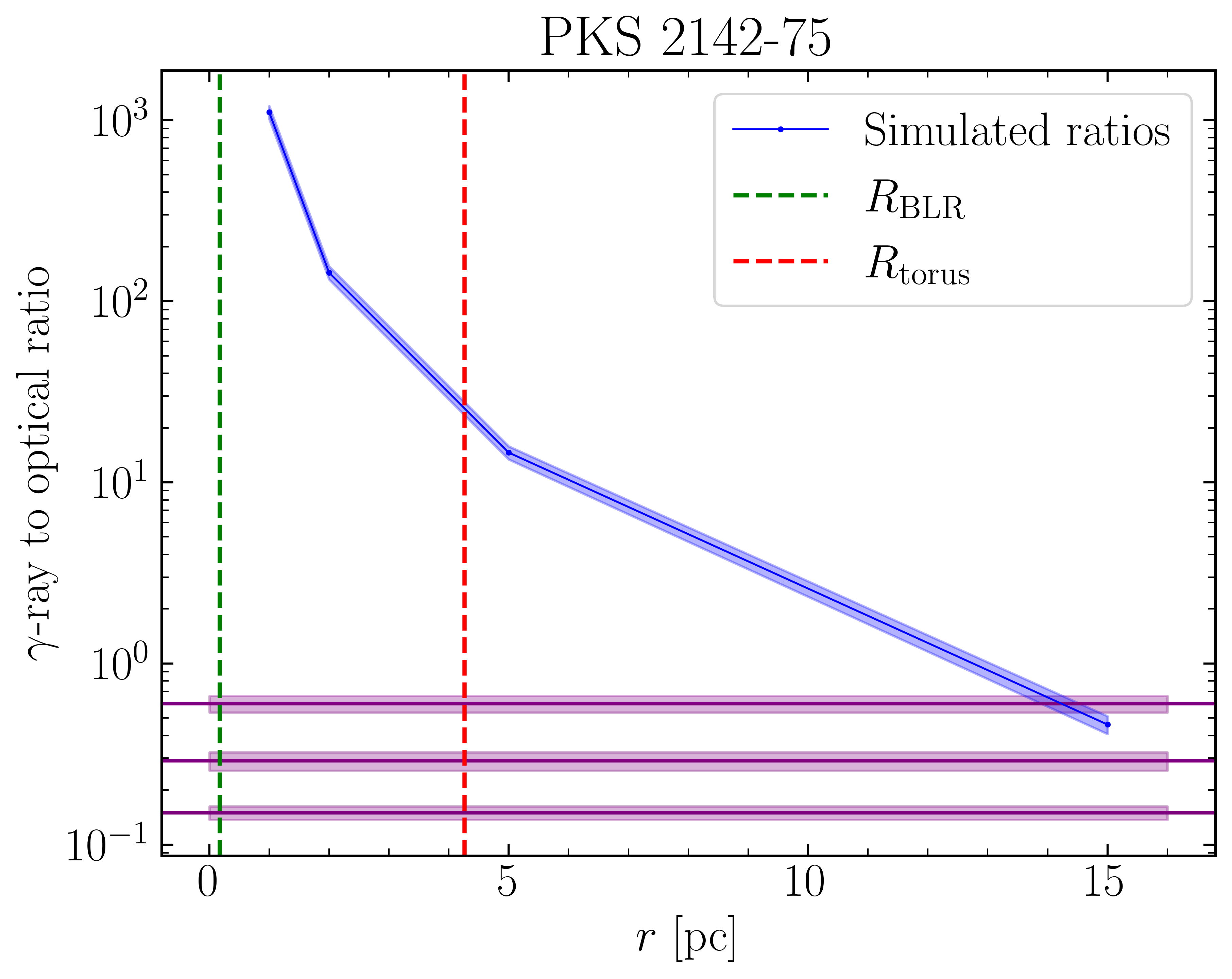}
        \includegraphics[scale = 0.4]{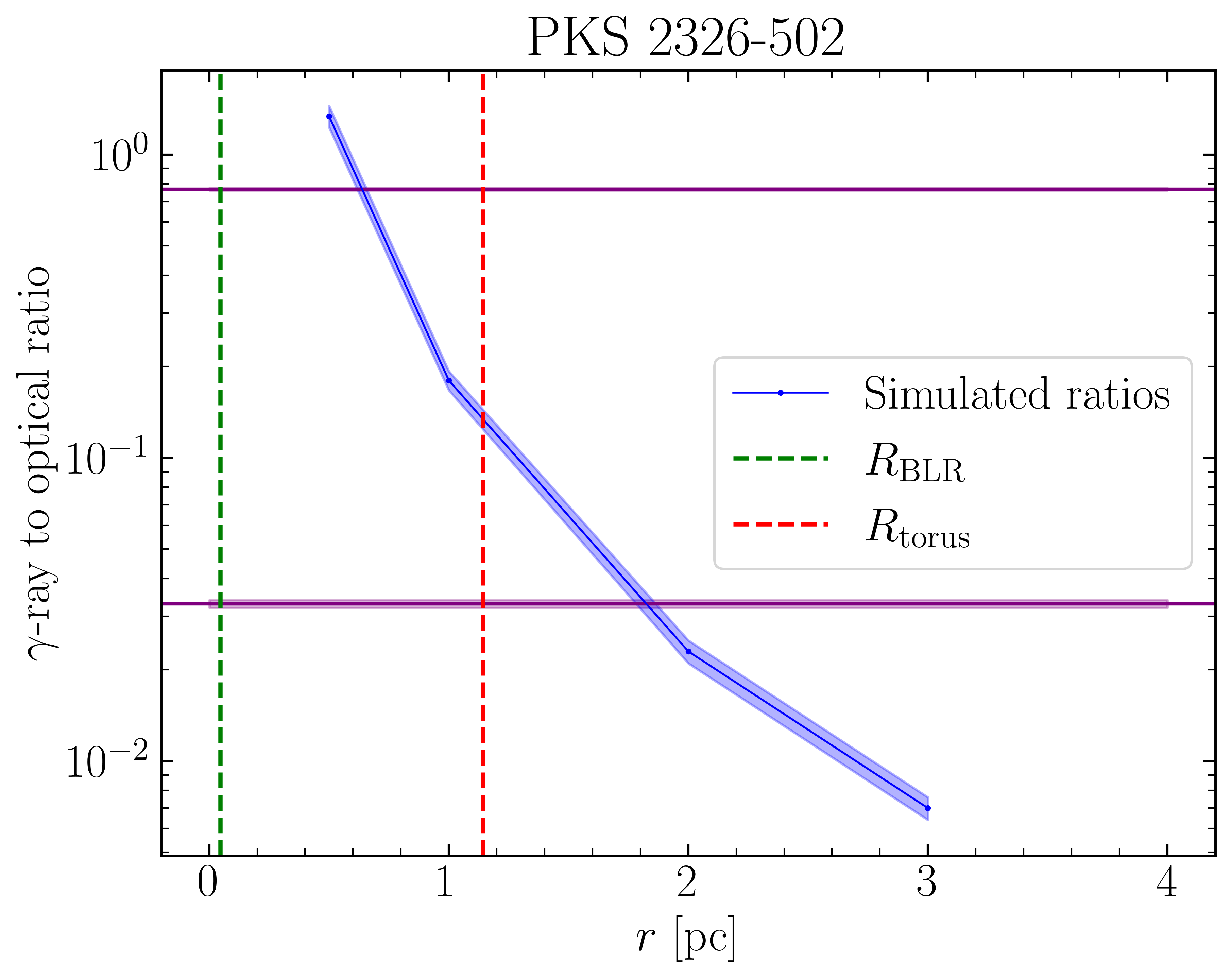}
    \caption{(continued)}
\end{figure*}

\begin{table*}
    \centering
    \begin{tabular}{c|c|c|c|c|c}
    \hline
    \hline
    Blazar Name & Location of BLR, & Location of torus, & Simulation & Simulated & Observed \\
    & $R_{\rm BLR}$ [pc] & $R_{\rm torus}$ [pc] & Distances [pc] & ratios & ratios \\
    (1) & (2) & (3) & (4) & (5) & (6)\\
    \hline
    \hline
        PKS 0208-512 & {0.06} & {1.58} & {0.25} & $9.13 \pm 0.87$ & $1.15 \pm 0.08$\\
        & & & {0.50} & $1.17 \pm 0.20$ & $0.26 \pm 9.04 \times 10^{-4}$ \\
        & & & {0.75} & $0.39 \pm 0.03$ & $0.29 \pm 0.001$ \\
        & & & {1.00} & $0.18 \pm 0.01$ & \\
    \hline
        PKS 0402-362 & 0.22 & 5.59 & {0.50} & $951.33 \pm 78.42$ & $0.51 \pm 0.06$ \\
        & & & {1.00} & $136.74 \pm 17.68$ & $2.36 \pm 0.21$\\
        & & & {2.50} & $1.16 \pm 0.45$ & $51.24 \pm 4.92$ \\
        & & & {3.50} & $0.29 \pm 0.01$ & \\
    \hline
        PKS 0426-380 & {0.1} & {2.5} & {0.50} & $2.65 \pm 0.25$ & $0.13 \pm 0.01$ \\
        & & & {1.00} & $0.32 \pm 0.02$ & $0.46 \pm 0.12$ \\
        & & & {1.25} & $0.17 \pm 0.02$ & $0.02 \pm 5.21 \times 10^{-3}$ \\
        & & & {1.50} & $0.11 \pm 0.01$ &  \\
    \hline
        PKS 0454-234 & {0.04} & {1.12} & {0.25} & $1.09 \pm 0.11$ & $0.82 \pm 3.40 \times 10^{-3}$ \\
        & & & {0.50} & $0.14 \pm 0.01$ & $0.06 \pm 1.03 \times 10^{-3}$ \\
        & & & {1.00} & $0.02 \pm 2.01 \times 10^{-3}$ & \\
    \hline
    \end{tabular}
    \caption{Results obtained using observed data and the simulated light curves. (1) Name of the target blazar. (2) Location of the BLR, estimated from $R_{\rm BLR} = 0.1 (L_{D, 46})^{0.5}~{\rm pc}$, where $L_{D, 46}$ denotes the disk-luminosity of the blazar in units of $10^{46}~{\rm erg}~{\rm s}^{-1}$. (3) Location of the torus, estimated from $R_{\rm torus} = 2.5 (L_{D, 46})^{0.5}~{\rm pc}$. (4) Distances of the $\gamma$-ray emission zone from the central engine, at which the simulation was run. (5) The simulated $\gamma$-to-optical ratios, averaged over all pairs of flares, obtained at the corresponding distance of column-4. (6) The $\gamma$-to-optical ratio for \textit{each flare-pair} of the relevant blazar. These values do not correspond to the distances shown in column-4.}
    \label{tab:blazar_results}
\end{table*}

\begin{table*}[]
    \ContinuedFloat
    \centering
    \begin{tabular}{c|c|c|c|c|c}
    \hline
    \hline
        Blazar Name & Location of BLR, & Location of torus, & Simulation & Simulated & Observed \\
        & $R_{\rm BLR}$ [pc] & $R_{\rm torus}$ [pc] & Distances [pc] & ratios & ratios \\
        (1) & (2) & (3) & (4) & (5) & (6) \\
    \hline \hline
        PKS 1244-255 & {0.09} & {2.23} & {0.50} & $0.86 \pm 0.07$ & $0.54 \pm 1.88$ \\
        & & & {1.00} & $0.11 \pm 0.01$ & $0.35 \pm 6.90$ \\
        & & & {2.00} & $0.02 \pm 1.34 \times 10^{-3}$ & $0.11 \pm 1.98 \times 10^{-3}$ \\
        & & & {3.00} & $7.59 \times 10^{-3} \pm 6.82 \times 10^{-4}$ & $0.01 \pm 0.04$ \\
        & & & & & $8.43 \times 10^{-1}$ \\
    \hline
        PKS 1510-089 & 0.06 & 1.62 & 0.25 & $472.08 \pm 53.46$ & $5.33 \pm 0.50$ \\
        & & & 0.50 & $134.50 \pm 15.19$  & $1.45 \pm 0.16$ \\
        & & & 1.00 & $64.57 \pm 5.52$ & $0.83 \pm 0.08$ \\
        & & & 1.50 & $14.24 \pm 1.26$ & $2.55 \pm 0.30$ \\
        & & & 2.00 & $4.18 \pm 0.36$ & $6.41 \pm 0.65$ \\
        & & & 2.50 & $1.49 \pm 0.17$ & $0.14 \pm 0.02$ \\
        & & & 3.00 & $0.33 \pm 0.04$ & $0.12 \pm 0.01$ \\
    \hline
        PKS 2142-75 & 0.17 & 4.26 & {1} & $1127.54 \pm 91.27$ & $0.56 \pm 0.06$ \\
        & & & {2} & $141.38 \pm 11.97$ & $0.32 \pm 0.03$ \\
        & & & {5} & $15.16 \pm 1.21$ & $0.13 \pm 0.01$ \\
        & & & {15} & $0.51 \pm 0.05$ & \\
    \hline
        3C 454.3 & {0.14} & {3.53} & {0.50} & $11.11 \pm 1.06$ & $1.96 \pm 0.03$ \\
        & & & {0.75} & $3.39 \pm 0.35$ & $3.43 \pm 0.02$ \\
        & & & {1.00} & $1.45 \pm 0.13$ & $0.82 \pm 0.01$ \\
        & & & {2.00} & $0.19 \pm 0.02$ & \\
    \hline
        PKS 2326-502 & {0.05} & {1.14} & {0.5} & $1.34 \pm 0.10$ & $0.77 \pm 7.05 \times 10^{-2}$ \\
        & & & {1.0} & $0.18 \pm 0.01$ & $0.03 \pm 1.01 \times 10^{-3}$ \\
        & & & {2.0} & $0.02 \pm 2.47 \times 10^{-3}$ & \\
        & & & {3.0} & $7.22 \times 10^{-3} \pm 6.09 \times 10^{-4}$ & \\
    \hline
        3C 273 & 0.219 & 5.477 & {1.0} & $11.23 \pm 1.08$ & $0.12 \pm 0.01$ \\
        & & & {1.5} & $4.77 \pm 0.43$ & $1.10 \pm 0.12$ \\
        & & & {2.0} & $1.26 \pm 0.11$ & $0.23 \pm 0.03$ \\
        & & & {3.0} & $0.08 \pm 6.18 \times 10^{-3}$ & $0.68 \pm 0.07$ \\
    \hline
    \end{tabular}

    \caption{(continued)}
\end{table*}

As per this procedure, the results obtained from the analysis of observed and simulated data are presented in Table \ref{tab:blazar_results}, and also depicted in Figure \ref{fig:ratio_plots}. We identify the following GeV emission zones for all the pairs of flares considered in the sample:
\begin{itemize}
    \item \textit{PKS 0208-512.} Three flare-pairs are found observationally for this source. All of them occur within the BLR and the torus, and {they are approximately midway between the BLR and the torus}. We see that all flaring activity takes place within \textbf{$\sim 0.9~{\rm pc}$} from the SMBH; the flare-pair originating closest to the SMBH is at a distance of \textbf{$\sim 0.5~{\rm pc}$}.

    \item \textit{{PKS 0402-362.}} {We have observed 3 flare-pairs for this source, for which the emission regions lie at various distances within the BLR and the torus. The closest flaring location to the central SMBH is at a distance of $\sim 1.4~{\rm pc}$, while the farthest is at $\sim 3.2~{\rm pc}$. }

    \item \textit{{PKS 0426-380.}} {
    We have extracted three flare pairs from the observed light curves of this source. For all of them, the emitting regions are between the BLR and the torus. The lowest $\gamma$-ray--to--optical ratio we have obtained from our simulations is $\sim 0.1$, at 1.5 pc. This is still higher than the lowest observed ratio, which implies that the upper limit of the emission region is between 2--2.5 pc.
    }

    \item \textit{PKS 0454-234.} We have two pairs of flares in this case, and all are found to be between the BLR and the torus. We have the upper limit to the distance of the flaring activity at $\sim 0.65~{\rm pc}$.

    \item \textit{PKS 2326-502.} {
    In this source, flaring does take place beyond the torus (at a distance of $\sim 2~{\rm pc}$, however, one pair originates within the torus as well.
    } The lower end of the flaring region is $\sim 0.6$ pc, roughly halfway between the torus and the BLR. 

    \item \textit{PKS 1510-089.} For this source, {the entire} outburst activity occurs beyond the torus, and it is one of the few sources in our sample to display {flaring beyond the torus}. The most upstream outburst is at a distance of $\sim 1.8~{\rm pc}$. The farthest flarings occur {beyond 3 pc}, as our simulated GeV-to-optical ratio at that distance is still higher than the lowest observed ratios (see Figure \ref{fig:ratio_plots}).

    \item \textit{{PKS 2142-75.}} {This is another source with all outbursts happening beyond the torus. We have observed 3 flare-pairs in this case, with GeV-to-optical ratios $\sim 10^{-1}-1$. The closest outbursts occur at $\sim 14.5~{\rm pc}$, while other flares even beyond 15 pc (see Figure \ref{fig:ratio_plots}). This is the only source in our samples with flaring activity at such large distances.}

    \item \textit{PKS 1244-255.} We have found 3 flare-pairs for this blazar. {
    While the uncertainties in the estimates of the observed $\gamma$-ray--to--optical ratios are relatively large compared to the other sources, we can constraint two flare pairs to originate between the BLR and the torus. The uncertainty on the last flare pair makes it difficult to determine its location with relative precision. In any case, it originates close to the torus, in a range of $\sim 2-2.8~{\rm pc}$. Thus, it may be slighly within or slightly beyond the torus.
    }

    \item \textit{3C 454.3.} We observe {3} flare-pairs for this blazar. {All of them} originate within the torus. {
    Two pairs are relatively close to the BLR, at $\sim 0.8-1$ pc, one pair is farther out, at $\sim 2~{\rm pc}$, roughly midway between the torus and the BLR.
    }

    \item \textit{3C 273.} Four flare-pairs were observed in 3C 273. It was seen that all four of them were generated between the BLR and the torus, at $\sim$ 2--3 pc.
\end{itemize}

Thus, from the above analysis, we see that in general the GeV Emission Region for LSP blazars lie beyond the broad line region. In a few cases, it may also lie beyond the torus. In most cases we have observed the emission region to be between the BLR and the torus.

\subsection{The Emitting Region for 3C 279}
\label{subsec:3C279}

In order to compare the constraint we obtain for the location of a $\gamma$-ray emitting region in a blazar jet with that determined by other authors using a different method for the same epoch of $\gamma$-ray outburst, we take the case of the blazar 3C 279. It is a well studied blazar at a redshift of $z = 0.536$ \citep{Burbidge1965}, and focus on one of its particular flaring episodes, observed in 2013--2014. This period of outburst has been studied extensively by \citet{Rani2018}. In their study, they found 6 short timescale $\gamma$-ray flares, superimposed on a longer timescale $\gamma$-ray outburst. Four of these outbursts also have optical and radio counterparts, as observed with \textit{Swift}-UVOT \citep{Roming2005} and Sub-Millimeter Array \citep[SMA;][]{Gurwell2007}. However, it was seen that the structure of outburst seen at optical and radio bands was missing in X-rays \citep[\textit{Swift}-XRT observations;][]{Burrows2005}. Only two of the six $\gamma$-ray flares had X-ray counterparts. 43 GHz very long baseline interferometry (VLBI) monitoring \citep{Jorstad2017} showed that, ejection of one VLBI component was simultaneous with the first three $\gamma$-ray flares, suggesting this component to be a potential source of GeV emission. Another such component was ejected at the time of the latter three $\gamma$-ray flares.

Based on those observations, \citet{Rani2018} concluded that while the later set of flares originated closer to the central SMBH the first set originated further upstream. In this section, we aim to determine the relative positions of the emitting regions corresponding to the two sets of flares, and compare the results obtained using our formalism with the above conclusion.

\begin{figure}
    \centering
    \includegraphics[width = \columnwidth]{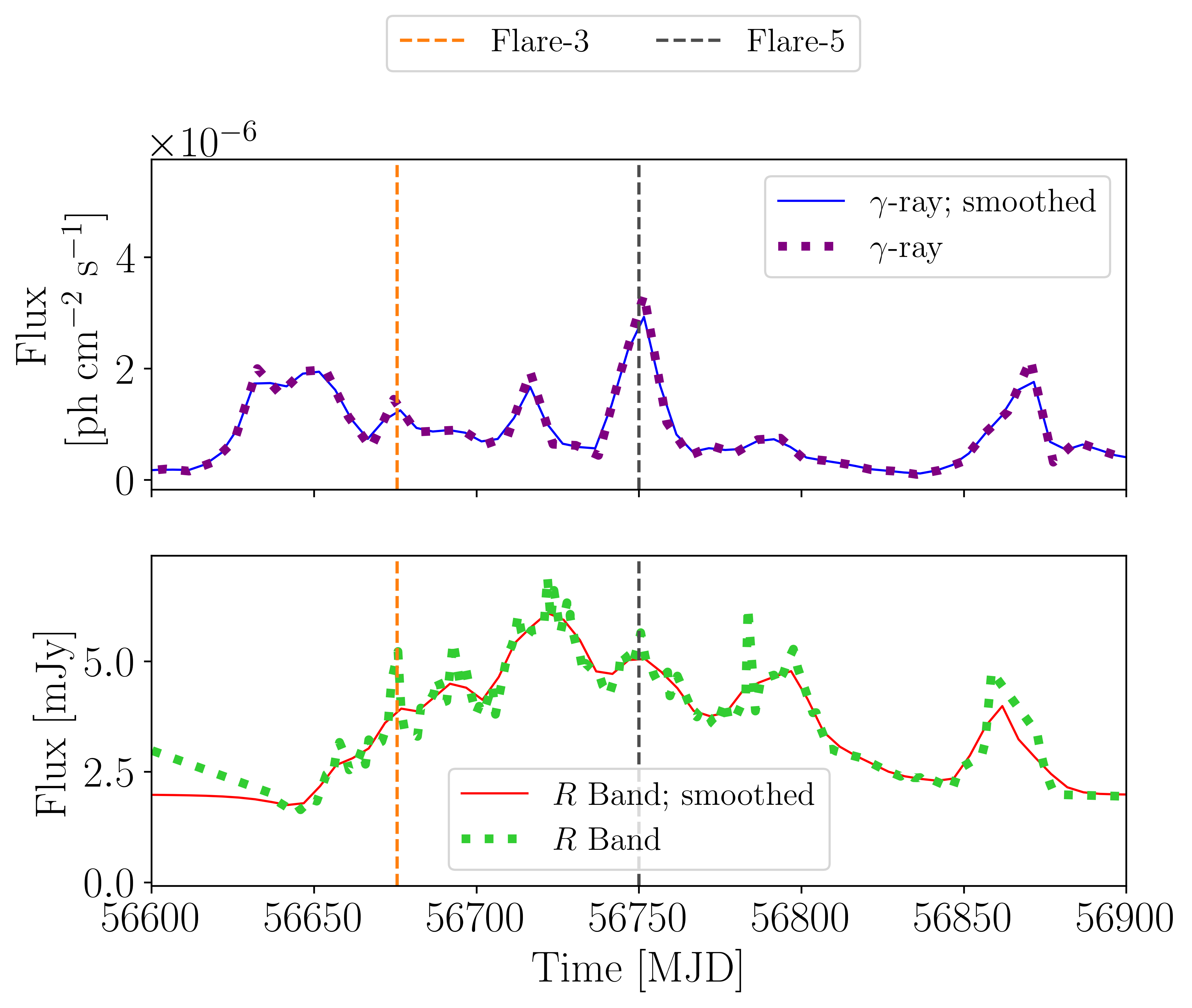}
    \caption{The light curves of 3C 279, during the 2013--2014 epoch of flaring: $\gamma$-ray \textit{(top)} and optical \textit{(bottom)}. The two flares, among the six flares in optical or $\gamma$-ray band are highlighted with dashed vertical lines (designated as `flare-3' and `flare-5'). The general flaring activity --- multiple flares superposed on a longer timescale outburst --- can clearly be seen.}
    \label{fig:3C279_Rani_LCs}
\end{figure}

In our analysis we focus on two of the six $\gamma$-ray/optical flares, namely, the third (designated as `flare-3') and fifth (designated as `flare-5') flares of the set, which are also discussed in details by \citet{Rani2018}. Flare-3 is thus simultaneous with the ejection of the first VLBI component along the jet, as flare-5 is with the second. As per the explanation of \citet{Rani2018}, they originate at different distances along the jet and finding the distances of their relative emitting regions would thus help to compare our results with those of \citet{Rani2018}.

From the light curves of Figure \ref{fig:3C279_Rani_LCs}, we determine the $\gamma$-to-optical energy ratios for flare-3 and flare-5. The ratios, in the units of ${\rm ph} \cdot {\rm cm}^{-2} \cdot {\rm s}^{-1} \cdot {\rm mJy}^{-1}$, are:
\begin{eqnarray}
    R_{\rm flare-3} &=& 7.15 \times 10^{-3} \label{eqn:R_flare3} \\
    R_{\rm flare-5} &=& 50.86. \label{eqn:R_flare5}
\end{eqnarray}

Thus, the $\gamma$-to-optical energy ratio is higher for the case of flare-5, than flare-3. This suggests that, the emitting region for flare-5 receives a relatively higher amount of seed photons due to its location closer to the central engine, than flare-3. This matches with the conclusions of \citet{Rani2018}.

Furthermore, we also analyze the relative positions of the two emitting regions, with respect to the BLR and the DT, by comparing the $\gamma$-ray and optical light curves with those generated from the simulation of the jet emission of 3C 279. The simulation parameters are shown in Table \ref{tab:3C279_parameters}. These parameters have mostly been obtained from the multi-wavelength analysis done by \citet{Roy2021}.

\begin{table}
    \centering
    \begin{tabular}{|c|c|}
       \hline
       \hline
       Parameter & Value \\
       \hline
       \hline
       $\Gamma$ & 20.9 \\
       $\gamma_{\rm min}$ & 1.1 \\
       $\gamma_{\rm max}$ & 4500 \\
       $B_i$ [G] & 1.0 \\
       $B_f$ [G] & 0.8 \\
       $L_D$ $[{\rm erg} \cdot {\rm s}^{-1}]$ & $10^{45}$ \\
       $R_{\rm BLR}$ [pc] & 0.03 \\
       $R_{\rm DT}$ [pc] & 0.81 \\
       $\theta_v$ & $2.4^{\circ}$ \\
       Doppler factor, $\delta$ & 23.67 \\
       \hline
    \end{tabular}
    \caption{The parameters used for simulating the model light curves of 3C 279.}
    \label{tab:3C279_parameters}
\end{table}

We have taken the disk luminosity of 3C 279 to be $L_D = 10^{45}~{\rm erg} \cdot {\rm s}^{-1}$, following \citet{Paliya2015}. The average magnetic field along the jet, in this case, is $\approx 0.9~{\rm G}$ \citep{Roy2021}. Hence, we set the magnetic field values at the base and the tip of the emitting region to be $B_i = 1.0~{\rm G}$ and $B_f = 0.8~{\rm G}$, respectively. The viewing angle to the blazar, and the bulk Lorentz factor of the jet, are fixed at $\theta_v = 2.4^{\circ}$ and $\Gamma = 20.9$, respectively \citep{Hovatta2009}. The Doppler factor, $\delta$, has been calculated as, $\delta = \{\Gamma (1 - \beta \cos\theta_v)\}^{-1}$.

\begin{figure}
    \centering
    \includegraphics[width = \columnwidth]{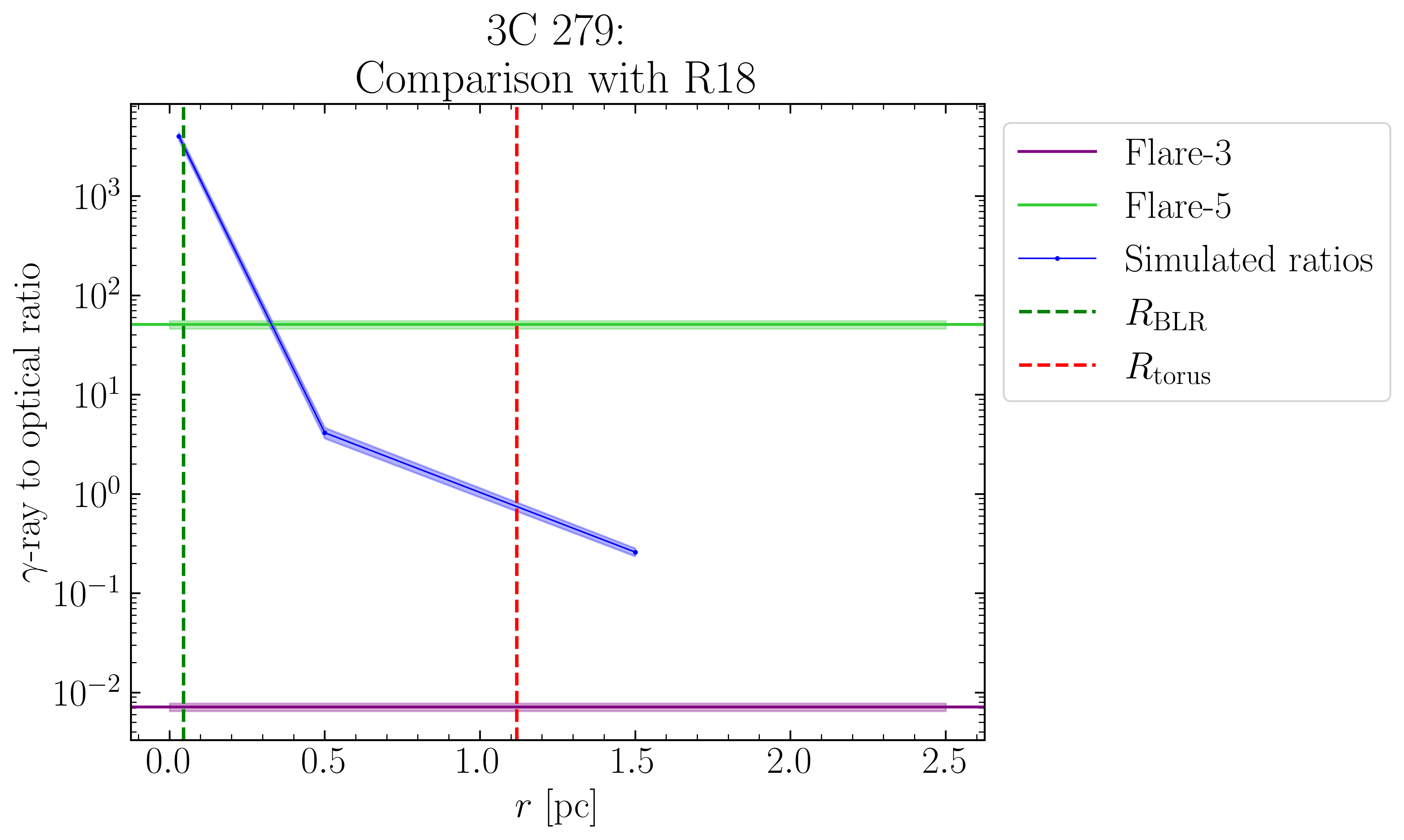}
    \caption{Variation of the $\gamma$-to-optical flare pair energy ratio with the distance of the emitting region from the SMBH. The dashed horizontal lines show the ratios for flare-3 and flare-5. The vertical lines denote the position of the BLR and the DT. The simulation is run at three different distances. The shaded regions correspond to relevant uncertainties. See text for details.}
    \label{fig:3C279_GTO}
\end{figure}

The simulation for 3C 279 is run at three different distances: 0.03 pc, 0.5 pc, and 1 pc. The flare pair energy ratios obtained at these distances are respectively 4000, 4.14, and 0.26, respectively, which are shown in Figure \ref{fig:3C279_GTO}. From Figure \ref{fig:3C279_GTO}, we see that the curve for the simulated flare pair energy ratios intersects the energy ratio value of flare-5 at a distance roughly mid-way between the BLR and the DT. Thus, flare-5 is generated beyond the BLR but within the torus.

However, we see that the energy ratio for flare-3 is not intersected by the simulated curve at even the farthest distance used in our simulation, which is 1 pc. We may safely assume that the simulated $\gamma$-to-optical energy ratio would decrease monotonically as we move past the DT, as there are no more seed photon sources to drive the GeV flares. Thus we can say that, at some distance even farther than 1 pc, the $\gamma$-to-optical ratio would be low enough that it would match that of flare-3. Since the torus is placed at $\approx 0.81~{\rm pc}$, we conclude that flare-3 originates even beyond the DT.

\subsection{Comparison with other existing studies}
\label{subsec:comparison}
We compare some of our obtained results, with similar studies which have also constrained the GeV emission zone in jets. \citet{Acharyya2021} --- hereafter A21 --- have constrained such emitting regions of several FSRQs by probing the variability and cooling timescales. Some of the sources in our sample are also included in their sample, in particular, 3C 454.3 and PKS 1510-089.

\subsubsection{3C 454.3}
Analyzing the December 2009 flare of 3C 454.3, A21 conclude that there may be multiple active GeV emission zones along the jet. Similar results have also been obtained by \citet{Pacciani2010}. We have obtained two emission zones, both of which are between the BLR and the torus. A21 have determined the emission zones to be both within the BLR and the DT. However, \citet{Coogan2016} have analyzed the December 2014 flare of 3C 454.3 to conclude that any BLR origin of the $\gamma$-ray emission is unlikely. Thus, the emission zone is likely to be between the BLR and the torus, which agrees with our results.

\subsubsection{PKS 1510-089}
{A21 argue that the GeV emission has a BLR origin from the November 2011 and February 2012 flares of PKS 1510--089. On the other hand, we have inferred multiple emitting regions outside the torus. However, A21 also suggested that the energy dependence of the cooling timescale indicates seed photons supplied by the DT. The detection of very high energy ($E \geq 0.1~{\rm TeV}$) photons from this source by the H.E.S.S. array of telescopes \citep{HESS2013} implies a low level of $\gamma$-$\gamma$ absorption, which makes the emitting region unlikely to be within the BLR \citep{Liu2006, Poutanen2010}. Using hour-scale variability, \citet{Brown2013} report that multiple emission zones may be present along the jet. We do not find any prominent GeV emission from within the BLR, in agreement with the observations of \citet{HESS2013}}.

\subsubsection{PKS 0208-512}
From the observations of the 2019--2020 flare of PKS 0208-512, \citet{Khatoon2022} used modeling of the broadband SED and particularly $\gamma$-ray spectra to estimate its $\gamma$-ray emitting region to be at $\sim$ 0.17 pc from the central SMBH. We obtain several emitting regions; most of them are situated within $\sim$ 0.5--1 pc. Thus our results agree with that of \citet{Khatoon2022} within a factor of a few. 
\citet{Stacy2003} have inferred the GeV emission region to be at $\sim$ 0.03 pc based on 4-day 
timescale of variability for this source. We have not found the GeV emission zone to be within the BLR for any of its flare pairs. However, location of the emission region directly inferred from the variability timescale involves an assumption that the entire jet cross-section is taking part in the variability, which is not necessarily true \citep[\textit{e.g.},][]{Hada2012,marscher14}.

\subsubsection{PKS 0454-234}
By determining the energy cut-off needed to explain VHE emission from PKS 0454-234, A21 have constrained the emitting region to be beyond the BLR. Similar results have also been obtained by \citet{Pacciani2014} by spectral analysis of the source. Our results support these findings; we have found the GeV emission zone at $\sim 0.5-1~{\rm pc}$ for this source, which is outside the BLR, but within the torus.

\subsubsection{3C 279}
We have found agreement with the results of \citet{Rani2018} for its November 2013--August 2014 flaring state. However, A21 have probed its variability timescale and found the GeV emission to come from a BLR origin. A similar result have been obtained by \citet{Hayashida2015} using the same method. In contrast, we have found two emission regions, one between the BLR and the torus, and another beyond the torus. However, the December 2015 flare was studied in the VHE domain by \citet{HESS2019} and the emission region was found to be beyond the BLR. That agrees with our results, which however corresponds to a different flaring state. A comparison with the exact same outburst, as in A21, was not possible due to lack of suitable GeV/optical data. {It is possible that the mismatch with the conclusion of A21 is due to the fact that their conclusion is based on one particular flare or the assumption that the entire jet cross-section is taking part in the variability, which may not be accurate.}

\subsubsection{3C 273}
The spectral variability study by \citet{Rani2013} determined the GeV emission location of 3C 273 to be within 1.6 pc from the central SMBH. A similar result was obtained by \citet{Chidiac2016}. By probing the correlation between the $\gamma$-ray and radio variability, they concluded the emission region to be $1.2 \pm 0.9~{\rm pc}$ from the apex of the jet although we must note that the central engine and the apex of the jet may not be coincident. However, for a strong flaring episode in September 2009, \citet{Lisakov2017} estimated the location of $\gamma$-ray emission to be at a distance of $2-7$ pc from the 43 GHz radio core. From broad line and jet emission variations, \citet{Liu2015} have estimated the emission zone to be close to the BLR. Our analysis reveals the $\gamma$-ray emission to be constrained within $2-3$ pc from the central engine, which is within a factor of a few of the distances found by other methods.

\subsubsection{PKS 0402-362}
{\citet{Das2023} found that the emitting region is located at the `edge' of the BLR while our analysis shows that the emitting regions can be within a pc from the BLR. Hence, our results are partially consistent with \citet{Das2023}.}


\subsubsection{\bf Limitations of the Method and Possible Causes of Mismatch with Other Studies}

{
From the above comparisons we notice that the results from our method are, in most cases, consistent with those obtained by applying different methods by other authors. However, there are some cases for which the results disagree. Below, we elaborate on certain aspects of other methods and the limitations of our method, which may lead to such mismatches. }

{ In this work, we use certain blazar parameters, listed in Table \ref{tab:blazar_parameters}, as input to our model in order to simulate GeV and optical light curves, which we then use to constrain the location of the emission region. Those parameters are different for each blazar and are obtained from published results. Naturally, the accuracy and precision of our results depend on those of the input parameters. In addition, our model makes certain assumptions about the geomtery of the BLR and torus with respect to the jet as well as about the emisison mechanisms involved. Sensitivity of the GeV/optical energy dissiptaion ratio to the location of the emission region depnds on the above factors. Furthermore, we can only apply our method to LSP blazars, for which the 0.1-100 GeV emission is dominated by the EC process. For time intervals, in which GeV/optical data of sufficient cadence are not available, the fits we perform to extract our results (see \S\ref{subsec:analysisObs}) may have larger uncertainties, e.g., in PKS 1244-255 (see Figure \ref{fig:ratio_plots}). This may lead to mismatch with other results. The above effects have been further discussed, quantitatively in some cases, in \S\ref{sec:discussion}. }

{ Different flares in a blazar can originate at different locations \citep[e.g.,][]{Acharyya2021, Brown2013, Barat2022}. Hence, our results may not match with those by other authors if different episodes are being compared. However, in some cases as discussed earlier in this section, we do find mismatch with other results for the same episode. Those may be due to the above limitations of our method. On the other hand, different methods used by oher authors also involve certain assumptions and difficulties, which may lead to mismatch with our results as well. For example, best-fit parameters obtained from modeling the SED of blazars often contain degeneracies. In order to break the degeneracy, certain assumptions are made, e.g., significant contribution from BLR seed photons to generate strong $\gamma$-ray flares, which favors an emission region closer to the central engine. In another example, A21 use shortest timescale of variability, cut-off or curvature of the GeV spectrum, TeV detection to infer their location. However, conversion of the variability timescale to a characteristic size of the emisison region requires an assumption of the Doppler factor, which contains significant uncertainty for most sources. Connecting the size to the location of the emission region involves the assumtion that the latter fillls up the entire jet cross-section, which may not be accurate (they mentioned this point as a caveat). They find the locations to be systematically closer to the central engine than those obtained by, \textit{e.g.}, \citet{Meyer_2019}. 
On the other hand, they find a curvature in the GeV spectrum of some of the blazars in their sample, which is consistent with being affected by the $\gamma$-$\gamma$ absorption in the BLR although, as mentioned by them, it does not confirm the location to be within the BLR. Furthermore, detections of TeV photons from blazars are not obtained by uniform all-sky surveys and hence it is not simple to connect those with contemporaneous GeV flares in order to infer about their locations. A21 also analyze the energy dependence of cooling timescales of the GeV flares to further verify the results obtained from the above methods. }

{ Since there is no umambiguous method to determine the exact location of GeV/optical outbrusts in the jets of blazars, application of different methods and verification of the consistency of the obtained results is a necessary approach, which has been employed in this work as well as by other authors. }




\subsection{Shoter-Timescale Flares}
\label{subsec:timescale}


We have so far focused on the longer (weeks to months) timescale outbursts, which are mainly driven by the injection of energetic particles in the emission region supposedly by a moving shock front. However, blazar light curves also contain shorter (hours to days) timescale variations. In our model, those smaller-amplitude shorter-timescale variability is primarily due to fluctuations of the magentic field in the emission region.

The GeV-to-optical ratio of these two classes of flares are expected to be different as well. For short timescale outbursts, the rise and decay times are short and hence the energy emitted for each GeV flare is smaller. For FSRQ-type blazars, the GeV emission is mostly due to the EC process, which is not affected by the small-scale fluctuations of the magnetic field. The optical emission, on the other hand, is due to synchrotron radiation which depends on the magnetic field: both the overall smoothly varying field and the small scale fluctuations.

Thus, for FSRQs, the ratio of the GeV to optical emission, which is a proxy for the EC to synchrotron radiation, depends on the ratio of the effect of change in the leptonic population due to shocks, to that due to overall variation (smoothly varying and turbulent) of the magnetic field.

Suppose the ratio of (GeV Emission) / (optical emission) be $R_1$ when both long and short timescale flares are considered, i.e., keeping $t_{\rm smooth}$ to a low value in the flare decomposition), and let it be $R_2$ when only longer timescale flares are considered (keeping $t_{\rm smooth}$ to a higher value). Between the first and the second case, the denominator (optical emission) is clearly larger for the first case, when both kinds of flares are included. Thus, from theoretical considerations, we expect $R_1 < R_2$. This is reflected in the observational data and in the simulated light curves as well (Table \ref{tab:timescale_FSRQ}).

However, for a BL Lac object, both the GeV Emission (due to SSC process) and the optical emission (due to synchrotron radiation) depend similarly on the magnetic field. Thus we theoretically do not expect any change in the GeV-to-optical ratio whether small scale turbulences are included or not. Indeed, we do not observe any such systematic change either in the observed or the simulated light curves (see Table \ref{tab:timescale_BL Lac}).

\begin{table}[]
   \centering
    \begin{tabular}{|c|c|c|}
    \hline \hline
        \textbf{Source} & $R_1$ [1] & $R_2$ [2] \\
    \hline \hline
         \multicolumn{ 3}{|c|}{\textsc{Observed Light Curves}} \\
    \hline
        PKS 0208-512 & $1.34 \times 10^{-1}$ & $3.50 \times 10^{-1}$ \\
        PKS 0454-234 & $2.55 \times 10^{-3}$ & $8.37 \times 10^{-2}$ \\
        3C 273 & $2.25 \times 10^{-1}$ & 2.10 \\
    \hline
        \multicolumn{ 3}{|c|}{\textsc{Simulated Light Curves}} \\
    \hline
        PKS 0208-512 & 0.66 & 2.15 \\
        PKS 0454-234 & 0.45 & 0.80 \\
        3C 273 & 1.32 & 3.54 \\
    \hline
    \end{tabular}
    \caption{GeV-to-optical ratios for some representative FSRQ Sources in our sample. Simulated Light Curves have been generated with parameters from \citet{Ghisellini2010}. [1] The ratio calculated considering both long and short timescale flares. [2] The ratio calculated with only long timescale flares.}
    \label{tab:timescale_FSRQ}
\end{table}

\begin{table}[]
    \centering
    \begin{tabular}{|c|c|c|}
    \hline \hline
        \textbf{Source} & $R_1$ [1] & $R_2$ [2] \\
    \hline \hline
         \multicolumn{ 3}{|c|}{\textsc{Observed Light Curves}} \\
    \hline
         OJ 287 & $3.37 \times 10^{-1}$ & $4.15 \times 10^{-1}$  \\
         PKS 2142-75 & 2.42 & 1.12 \\
         PKS 1424-75 & 1.20 & 1.19 \\
         0531-4827 & $4.50 \times 10^{-1}$ & $3.84 \times 10^{-1}$ \\
         PKS 1244-255 & 2.96 & 3.46 \\
    \hline
        \multicolumn{ 3}{|c|}{\textsc{Simulated Light Curves}} \\
    \hline
        OJ 287 & $1.18 \times 10^{-1}$ & $2.34 \times 10^{-1}$ \\
        PKS 2142-75 & 4.27 & 3.98 \\
        PKS 1424-75 & $1.46 \times 10^{-1}$ & $1.93 \times 10^{-1}$ \\
    \hline
    \end{tabular}
    \caption{GeV-to-optical ratios for some representative BL Lac Sources in our sample. Simulated Light Curves have been generated with parameters from \citet{Ghisellini2010}. [1] The ratio calculated considering both long and short timescale flares. [2] The ratio calculated with only long timescale flares.}
    \label{tab:timescale_BL Lac}
\end{table}


\section{Discussion}
\label{sec:discussion}

The location of the $\gamma$-ray emitting region in blazar jets has long been a matter of debate in the community because, in general, the jet is unresolved, except at radio wavelengths. However, determining the location is important in order to constrain the emission processes as well as geometric and physical parameters of the jet. In addition, the location has crucial effects on the nature of interaction of the jet with the interstellar and intergalactic medium \citep[\textit{e.g.},][]{Mukherjee2021}.

During the \textit{Fermi} era, tens of blazars have been regularly monitored at multiple wavelengths at days to years timescale. Various methods have been employed to constrain the exact location of the nonthermal emission in blazar jets using those data. Different kinds and amounts of data are necessary to apply those methods, \textit{e.g.}, modeling of the SED, shortest timescale of variability, detection of TeV $\gamma$-rays, etc.

\citet{Barat2022} established a method in which the location of the GeV/optical emission in blazar jets is determined using the ratio of energy emitted during contemporaneous outbursts at those wave bands. That method is applicable to mainly low synchrotron peaked (LSP) blazars, in which the GeV emission is dominated by the EC process while the optical emission is due to synchrotron radiation. The advantage of the method is that the required long-term GeV and optical light curves may be retrieved for many blazars in publicly available archives. On the other hand, the location information is obtained by comparing the the GeV/optical energy dissipation ratio of correlated multi-wavelength outbursts with that simulated by a theoretical model, which introduces certain dependence on the assumptions of the model.

Here, we have applied the method established by \citet{Barat2022} to a total of 47 multi-wavelength outbursts in ten blazars. The results obtained here demonstrate that the location of the GeV/optical emission in a statistically significant sample of blazars is away from the SMBH beyond the BLR for all cases. This result is consistent with other studies, in which the location has been determined for a large sample of blazars \citep[\textit{e.g.},][]{Harvey2020}. Furthermore, this work also provides us the opportunity to compare the location determined by us for a GeV flare of a blazar to that obtained by other authors using a different method for that exact episode. We find that our results are consistent in such one-to-one comparison in most cases, for which the required data were available.

To determine the location of the BLR and the DT in a given blazar, we use scaling relations, which connect the distance of the BLR and DT from the SMBH with the the accretion disk luminosity of the AGN. These relations are derived from reverberation mapping \citep[\textit{e.g.},][]{Bentz2013} and such estimates of the BLR and the DT locations do have uncertainties. Equations \ref{eqn:u_EC_BLR} and \ref{eqn:u_EC_DT} \citep{Hayashida2012} suggest that, for the cases where the emission region is within the BLR, or between the DT and the BLR, the seed photon field intensity may be taken to be essentially constant, and it decreases rapidly outside this region \citep{BottcherEls2016}. Thus, it is only the relative locations of the BLR and the DT that are important. We have used magnetic field values at the base ($B_i$) and the tip ($B_f$) of the emitting region based on average magnetic field values in the jets, from, \textit{e.g.}, {\citet{Paliya_2017}}. These values also have uncertainties. The value of the magnetic field is important as it sets the intensity of the optical synchrotron emission. We see that, for a particular parameter-set for a certain blazar, the average $\gamma$-to-optical ratio scales roughly with the magnetic field as $B^{-2}$ (this is expected as the emitted power of the synchrotron radiation scales as $B^2$). Thus, we see that, if the magnetic field is varied from its given value to, say, within a factor of 10, the emission region still remains in the same location relative to the BLR and the DT. However, it is seen that, varying, \textit{e.g.}, $L_D$ to 10 times of the values used, moves the GeV emission zone outwards for $\sim 2.5$ pc, on an average, which generally changes the location of the emission zone relative to the BLR and the DT, \textit{e.g.}, if it was between the BLR and the DT, it may move to beyond the DT. Similarly, an opposite trend (upstream movement of the emission zone) is expected if $L_D$ is decreased from its used values. 
Thus our method is sensitive to parameters which govern the EC emission (\textit{e.g.}, $L_D$, $\Gamma$).

The comparison with the 2013-14 flares in the blazar 3C 279 is of particular note because the location of those outbursts have been constrained precisely using a combination of detailed multi-wavelength light curves and VLBA monitoring of the pc-scale jet. Agreement with that result, as well as other outbursts in several other sources obtained by different techniques, using our method provides an indication that our assumptions are accurate at the level of precision provided by those obsevations. Furthermore, it provides stringent constraints on the geometrical properties and emission processes for jet emission models.

We find that even accounting for the uncertainties in our assumed parameter values the location of most or all of the GeV/optical outbursts are away from the SMBH a few pc down the jet between the BLR and torus or beyond. That emphasizes the importance of the seed photons from the torus for the production of GeV emission. It also implies the jet contains energetic electrons capable of producing the observed luminous outbursts far away from the central engine. The magnetic field, Doppler factor and other free parameters that are often determined using various properties of the GeV/optical emission are also properties of the jet a few pc from the SMBH. Those provide important constraints for theoretical modeling of the launching and collimation of blazar jets as well as their interaction with the surrounding medium.

{Traditionally, powerful GeV flares have been difficult to produce in SED models without significant contribution from the BLR photons. Consequently, such studies have usually concluded blazar outbursts to be produced within the BLR, e.g., $\sim 0.1-1$ pc from the central engine. However, this study and \citet{Harvey2020} demonstrate that a larger fraction of the outbursts may be generated few pc down the jet from the SMBH. Similar inferences have also been made by studies involving VLBA monitoring of pc-scale dynamics of the jet and its conection with multi-wavelength variability \citep[\textit{e.g.},][]{Marscher2010, Agudo2011}.}

We also found that the GeV/optical energy dissipation ratios for short-timescale flares in the blazars in our sample match with that from the model results. Short-timescale variability information at multiple wave bands is becoming available for larger samples of blazars in recent years. A comparison of those data with a more expanded version of our model will be useful to probe the small-amplitude short-tiemscale fluctuations in blazar emisison produced by, \textit{e.g.}, a turbulent magnetic field.

\section*{Acknowledgements}

{The authors wish to thank the anonymous referee whose comments and suggestions have greatly improved the draft.}
MK acknowledges financial support from {Washington University in St.\ Louis through the McDonnell Center Graduate Fellowship}.
MK thanks Aritra Kundu for help with the usage of the jet emission code, and thanks Samrat Roy for helpful discussions. We thank Kaustav Mitra for providing the older version of the jet emission code that was used in this work. {AB acknowledges financial support from the UGC-NET fellowship.} AB and RC thank IUCAA for their hospitality and usage of their facilities during their stay at different times as part of the university associateship program. RC thanks ISRO for support under the \textit{AstroSat} archival data utilization program, and thanks Presidency University for support under the Faculty Research and Professional Development (FRPDF) Grant. RC acknowledges financial support from Anusandhan National Research Foundation (ANRF) for a SERB-SURE grant (File No.\ SUR/2022/001503).

\section*{Data Availability}

We have used data from the following sources:
\begin{itemize}
    \item[--] \textbf{The \textit{Fermi}-LAT 10-year Source Catalog.} Link: \\
    \url{https://fermi.gsfc.nasa.gov/ssc/data/access/lat/10yr_catalog/}

    \item[--] \textbf{Yale-SMARTS Optical Blazar Light Curves.} Link: \\
    \url{astro.yale.edu/smarts/glast/home.php}
\end{itemize}

\bibliographystyle{mnras}
\bibliography{manuscript}

\bsp	
\label{lastpage}
\end{document}